\documentclass[sigconf]{acmart}
\AtBeginDocument{%
  \providecommand\BibTeX{{%
    \normalfont B\kern-0.5em{\scshape i\kern-0.25em b}\kern-0.8em\TeX}}}

\usepackage{graphicx}
\usepackage{amsmath}
\usepackage{amsfonts}
\usepackage{amsthm}
\usepackage{appendix}
\usepackage{subfigure}
\usepackage{caption}
\usepackage{bm}

\newtheorem{theorem}[]{Theorem}

\newcommand{\ev}[2]{\mathbb E_{#1} \left [ #2 \right ]}

\newcommand{\one}[0]{\bm {1}}

\newcommand{\til}[0]{\widetilde}

\newcommand{\var}[2]{\mathrm{Var}_{ #1 } \left [ #2 \right ]}

\newcommand{\manualnumber}[1]{\noindent \textbf{\emph{#1}}}

\newcommand{\argmax}{\mathrm{argmax}}
\newcommand{\argmin}{\mathrm{argmin}}

\newcommand{\pwerr}[0]{\mathsf {PWFitError}}

\newcommand{\funding}{Supported in part by a Simons Investigator Award, a Vannevar Bush Faculty Fellowship, AFOSR grant FA9550-19-1-0183, a Simons Collaboration grant, and a grant from the MacArthur Foundation.}
\newcommand{\acknowledgements}{The authors would like to thank A. Benson, E. Pierson, A. Singh, K. Tomlinson, J. Ugander, and Y. Wang for the useful discussions.}

\copyrightyear{2022}
\acmYear{2022}
\setcopyright{acmcopyright}\acmConference[KDD '22]{Proceedings of the 28th ACM
SIGKDD Conference on Knowledge Discovery and Data Mining}{August 14--18,
2022}{Washington, DC, USA}
\acmBooktitle{Proceedings of the 28th ACM SIGKDD Conference on Knowledge
Discovery and Data Mining (KDD '22), August 14--18, 2022, Washington, DC, USA}
\acmPrice{15.00}
\acmDOI{10.1145/3534678.3539272}
\acmISBN{978-1-4503-9385-0/22/08}

\begin{document}

\title{Core-periphery Models for Hypergraphs}

\author{Marios Papachristou}
\email{papachristoumarios@gmail.com}
\affiliation{
    \institution{Cornell University}
    \city{Ithaca}
    \country{United States}
}

\author{Jon Kleinberg}
\email{kleinberg@cornell.edu}
\affiliation{
    \institution{Cornell University}
    \city{Ithaca}
    \country{United States}
}

\renewcommand{\shortauthors}{Papachristou and Kleinberg}
\begin{abstract}
    We introduce a random hypergraph model for core-periphery structure. By leveraging our model's sufficient statistics, we develop a novel statistical inference algorithm that is able to scale to large hypergraphs with runtime that is practically linear wrt. the number of nodes in the graph after a preprocessing step that is almost linear in the number of hyperedges, as well as a scalable sampling algorithm. Our inference algorithm is capable of learning embeddings that correspond to the reputation (rank) of a node within the hypergraph. We also give theoretical bounds on the size of the core of hypergraphs generated by our model. We experiment with hypergraph data that range to $\sim 10^5$ hyperedges mined from the Microsoft Academic Graph, Stack Exchange, and GitHub and show that our model outperforms baselines wrt. producing good fits. 
\end{abstract}



\maketitle

\section{Introduction}

Numerous real-world higher-order networks exhibit the so-called core-periphery structure. In this structure, the network is  roughly comprised of a \emph{core} and a \emph{periphery}. The core is a set of nodes where the nodes are \emph{tightly connected} with one another, and almost cover the periphery of the network \cite{papachristou2021sublinear}. The nodes that lie within the periphery of the graph are \emph{sparsely connected} with one another and connected to the core. In terms of \emph{generative model} formulations \cite{zhang2015identification, elliott2020core,papachristou2021sublinear} study core-periphery structure on graphs and unravel interesting insights regarding the core of the network and its identification. Remarkably, the recent work of \cite{papachristou2021sublinear} observes that the core of (small-scale) real-world graphs is of \emph{sublinear} size wrt. the number of nodes, by using a random graph model that builds on a \emph{communities-within-communities}, i.e. fractal-like, \cite{leskovec2007graph, kleinberg2002small} skeleton where nodes are more likely to attach to other more prestigious nodes, exhibiting small-world phenomena, and power laws.

Core-periphery structure has a rich literature \cite{nemeth1985international, wallerstein1987world, zhang2015identification, snyder1979structural} and appears in a variety of contexts \cite{bonato2010geometric, bonato2012geometric, bonato2015domination, nacher2012dominating, nacher2013structural, rombach2017core, della2013profiling, boyd2010computing}. Moreover, a series of works have been devoted to identifying this structure on graphs \cite{jia2019random, tudisco2019fast, tudisco2019nonlinear, zhang2015identification, borgatti2000models} using generative graph models that provide us with useful insights regarding the understanding of graph structure \cite{leskovec2007graph, newman2003random}. More specifically, the work of \cite{jia2019random} provides a logistic model for core-periphery structure that is able to incorporate spatial characteristics and provides a \emph{core score} $z_i \in \mathbb R$ for every node $i$ in the graph which determines the distance of each node to the core and has a gradient evaluation cost of $O(n \log n + m)$ for a graph on $n$ nodes and $m$ edges. The models of \cite{tudisco2019fast, tudisco2019nonlinear} are based on inferring a permutation $\pi$ of the nodes based, again, on a logistic model of edge formation and propose iterative non-linear spectral algorithms which have $O(m)$ per-step complexity.

Moreover, everyday interactions are governed by \emph{higher-order} relationships, i.e. interactions between groups, such as e-mail threads, research collaborations, and QA forums, which yield combinatorial structures known as \emph{hypergraphs}. Despite the growing attention around hypergraphs, studying and \emph{tractably} recovering the core-periphery structure in hypergraphs has general been a relatively unresolved question. 

A generative model for core-periphery structure which can be compared with our model, though \emph{not as scalable as} our proposed model, is the generalization of the baseline logistic model of  \cite{jia2019random}, which is also known as the $\beta$-model \cite{wahlstrom2016beta,stasi2014beta}. Besides, the work of \cite{amburg2021planted} studies the recovery of the core in hypergraphs via the \emph{different} notion of ``core-fringe structure'' (see also \cite{benson2018found} for the graph analogue). The core-fringe structure can be summarized, in a nutshell, as: in an e-mail network of a company, addresses of people within the company make up the \emph{core} and addresses of people outside the company -- the which have exchanged emails with people in the company, but not with one another (at least in the measured network), make up the \emph{fringe}. In such case we are aware \emph{only} of the core-core and core-fringe links, and the fringe-fringe links are generally \emph{non-existent in the data}. Thus, \cite{amburg2021planted} models \emph{lack of information} in a network and attempts to recover a (large) core, whereas our work models \emph{core-periphery} structure in a \emph{continuous and hierarchical manner}. It is therefore clear that the two works \emph{cannot} be experimentally compared since they refer to different structural characterizations.

\emph{Concurrently} and \emph{independently} of our method, the work of \cite{tudisco2022core} examines core-periphery identification in hypergraphs based on a generative model -- HyperNSM -- and develops a non-linear spectral method to rank hypergraph nodes according to their prestige. The main differences with our work are: Firstly, in our work we can account for \emph{features} -- without compromising computational complexity -- on the nodes where their method assumes only hypergraph structure . Secondly, the log-likelihood of our model is \emph{tractable} whereas the HyperNSM log-likelihood is not generally tractable. Thirdly, we provide an efficient and principled method to \emph{sample} hypergraphs from our model, whereas it is generally intractable to sample hypergraphs from HyperNSM. Finally, we compare with HyperNSM and conclude that our model produces \emph{better fits} (in terms of the log-likelihood) with respect to HyperNSM.

\noindent \textbf{Contribution.} In this paper, we make the following contributions: 

\manualnumber{1.} We introduce -- concurrently and independently with \cite{tudisco2022core} -- the \emph{novel inference task} of identifying the (continuous) core-periphery structure of hypergraphs, and introduce a \emph{novel hierarchical model} for learning this structure in hypergraphs (Sec.~\ref{sec:model})

\manualnumber{2.} We develop a tractable \emph{exact inference} algorithm to recover the core-periphery structure of hypergraphs. We show that for the proposed models, and after careful preprocessing which is (almost) linear in the hypergraph order and the \emph{number of edges}, the graph likelihood can be computed in \emph{linear time} wrt. the \emph{number of nodes} for a constant number of layers (Sec.~\ref{sec:hyperedge_partitioning}, Sec.~\ref{sec:inference}). Moreover, when $d$-dimensional features are provided, we are able to extend this scheme to infer endogenous (latent) ranks by paying an $O(nd + m)$ cost-per-step (Sec.~\ref{sec:learning_ranks}). 

\manualnumber{3.} We develop an efficient \emph{sampling} algorithm (Sec.~\ref{sec:sampling}).

\manualnumber{4.} We prove that, under certain conditions, the size of the core (which is viewed as a dominating set) is \emph{sublinear} wrt. the size of the graph (Sec.~\ref{sec:small_core}). 

\manualnumber{5.} Experimentally (Sec.~\ref{sec:experiments}), we \emph{fit} our model to real-world networks of various sizes and disciplines. We compare our model with the logistic models of \cite{jia2019random,stasi2014beta,wahlstrom2016beta}, and HyperNSM \cite{tudisco2022core} and show that our model is able to produce better fits. 

\noindent \textbf{Source Code.} \cite{anonymous_2022_5965856}  \textbf{Data.}  \cite{anonymous_2022_5943044}

\section{Model} \label{sec:model}

The use of \emph{hierarchical} (or \emph{self-similar}, or \emph{fractal-like}) models for graphs has enabled the understanding and modelling of important graph phenomena such as densification laws \cite{leskovec2007graph}, core-periphery structure \cite{papachristou2021sublinear}, and are observed in a variety of graphs such as computer networks, patent networks, social networks \cite{watts2002identity,menczer2002growing,leskovec2010kronecker}. Briefly, models which enable us to describe self-similar structures usually obey the \emph{communities-within-communities} structure in terms of a finite-height $b$-ary tree which is then combined with a (random) edge-creation law to generate a random graph. 

As we noted, such models have been extensively and successfully used to model \emph{graphs}, yet the generation of \emph{hypergraphs} from such hierarchical models is a completely new task. Although hypergraphs are the obvious generalization of graphs, tasks regarding random hypergraph models pose new challenges, primarily from a computational standpoint, which correspond to the tractability of computing the respective log-likelihood function (see Sec.~\ref{sec:fast_algorithms}). Furthermore, the \emph{hybrid} nature of the existing models (i.e. involving both discrete and continuous structures) poses challenges wrt. the corresponding optimization problems of fitting the data. Here existing models such as \cite{leskovec2007graph,papachristou2021sublinear} use \emph{heuristic} methods to recover the model parameters given data, which in general do \emph{not} correspond to the maximum likelihood estimates.  

The proposed model, which we name the \emph{Continuous Influencer-Guided Attachment Model} (CIGAM) is a \emph{novel random hypergraph model} that overcomes the aforementioned challenges. Our model starts with a hypergraph $G(V=[n], E)$ on $n$ nodes where each node $i \in [n]$ is associated with a (probably learnable) \emph{prestige} value $r_i \in [0, 1]$ which we call the \emph{rank of node $i$} and is generated (i.i.d.) from a \emph{truncated exponential distribution}\footnote{It has been shown that many creative rank-based measures follow power-laws~\cite{clauset2009power}. The log-transforms of such quantities are exponentially-tailed.}, i.e.

\begin{equation} \label{eq:truncated_exponential}
    p(r_i) \propto \lambda e^{-\lambda r_i} \one \{ r_i \in [0, 1] \} \text { for } \lambda > 0.  
\end{equation}

We define $b = e^\lambda > 1$. For simplicity of exposition, we start with presenting the \emph{single-layer} CIGAM model. The model generates hyperedges of any order $2 \le k \le n$. Conceptually, we want the hyperedge creation probabilities to exhibit attachment towards more prestigious nodes; in agreement with existing generative core-periphery models \cite{jia2019random,tudisco2019nonlinear,stasi2014beta}. Subsequently, we want the edge creation law $f(e)$ to depend on $r_i$ for $i \in e$, and be scale-free \cite{leskovec2007graph,papachristou2021sublinear}.    

For the former property, we assume that $f(e)$ depends on $r_e = (r_i)_{i \in e}$, and specifically on $\| r_e \|_\infty = \max_{i \in e} r_i$. For the latter property, we want that for two edges $e, e'$ with $r_{e'} = r_e + \delta$ and for any appropriately chosen $\delta$ to obey $\frac {f(e')}{f(e)} = c^{\delta}$. This functional equation has a solution of the form $f(e) \propto c^{\| r(e) \|_\infty}$, and, thus, we generate each hyperedge $e$ independently with probability 

\begin{equation} \label{eq:single_layer_cigam} \tag{SL-CIGAM}
    f(e) = c^{-\zeta + \| r_e \|_\infty} 
\end{equation}
 
for $c, \zeta > 1$. We define $|e|$ to be the size of $e \in E$, $k_{\min} = \min_{e \in E} |e|$, and $k_{\max} = \max_{e \in E} |e|$ (also known as the \emph{hypergraph rank}). 

From \eqref{eq:truncated_exponential}, we observe that CIGAM overcomes the design challenge of a hybrid model since learning the parameters of CIGAM corresponds to solving a continuous optimization problem (see also Sec.~\ref{sec:inference}), and that the hyperedge creation probabilities of \eqref{eq:single_layer_cigam} are determined by the node with the highest prestige. Thus, the highest-prestige nodes serve as \emph{sufficient statistics} to \emph{simplify} the log-likelihood of the model. 

The previous definition can be extended to a multi-layer model parametrized by $\lambda > 0$, and $1 < c_2 \le c_3 \le \dots \le c_L$, $H_0 = 0 \le H_1 \le H_2 \dots \le H_L = 1$  as follows: We define a function $\phi: E \to [L]$ such that $\phi(e) = \inf \{ i \in [L] : 1 - \| r_e \|_{-\infty} \ge H_i \} $, where $\| r_e \|_{-\infty} = \min_{u \in e} r_u$, and draw each edge $e$ independently with probability 

\begin{equation} \label{eq:multi_layer_cigam} \tag{ML-CIGAM}
    f(e) = c_{\phi(e)}^{-\zeta+\| r_e \|_\infty}.
\end{equation}

\begin{figure}[t]
    \centering
    \includegraphics[width=0.4\columnwidth]{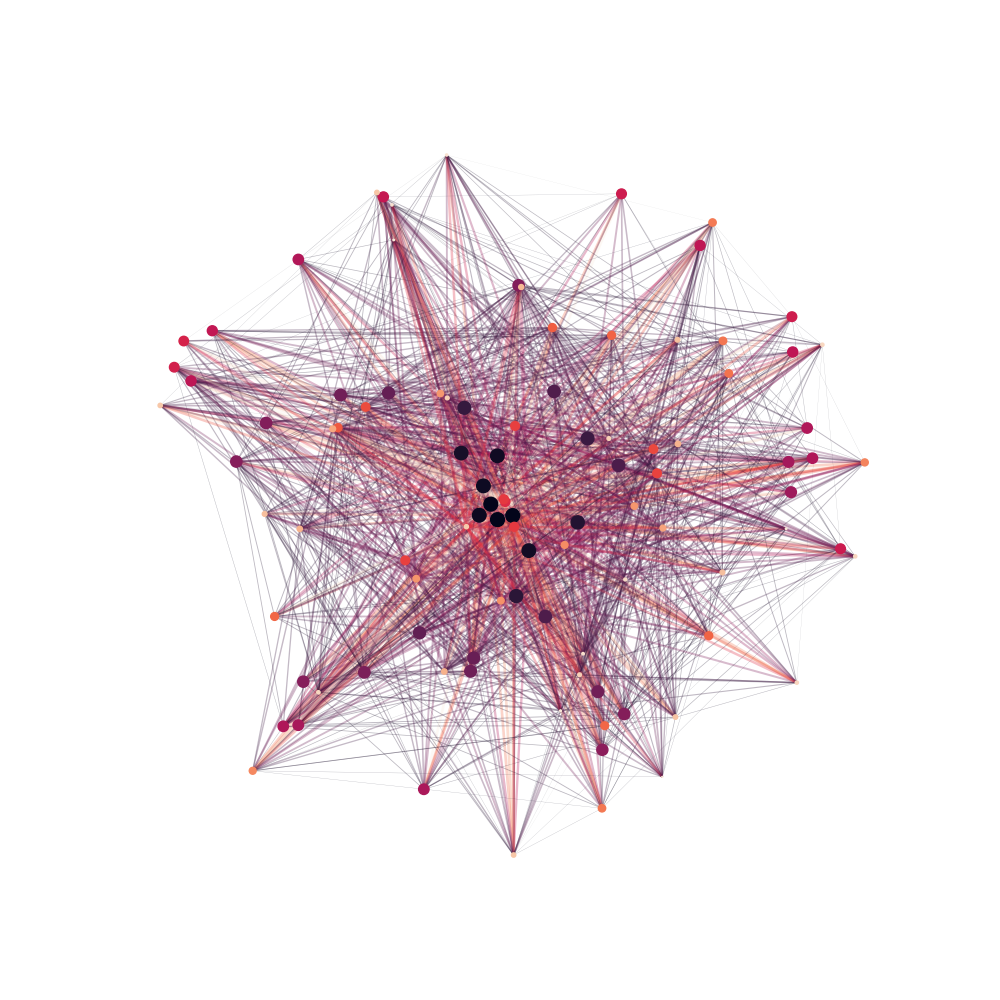}
    \includegraphics[width=0.4\columnwidth]{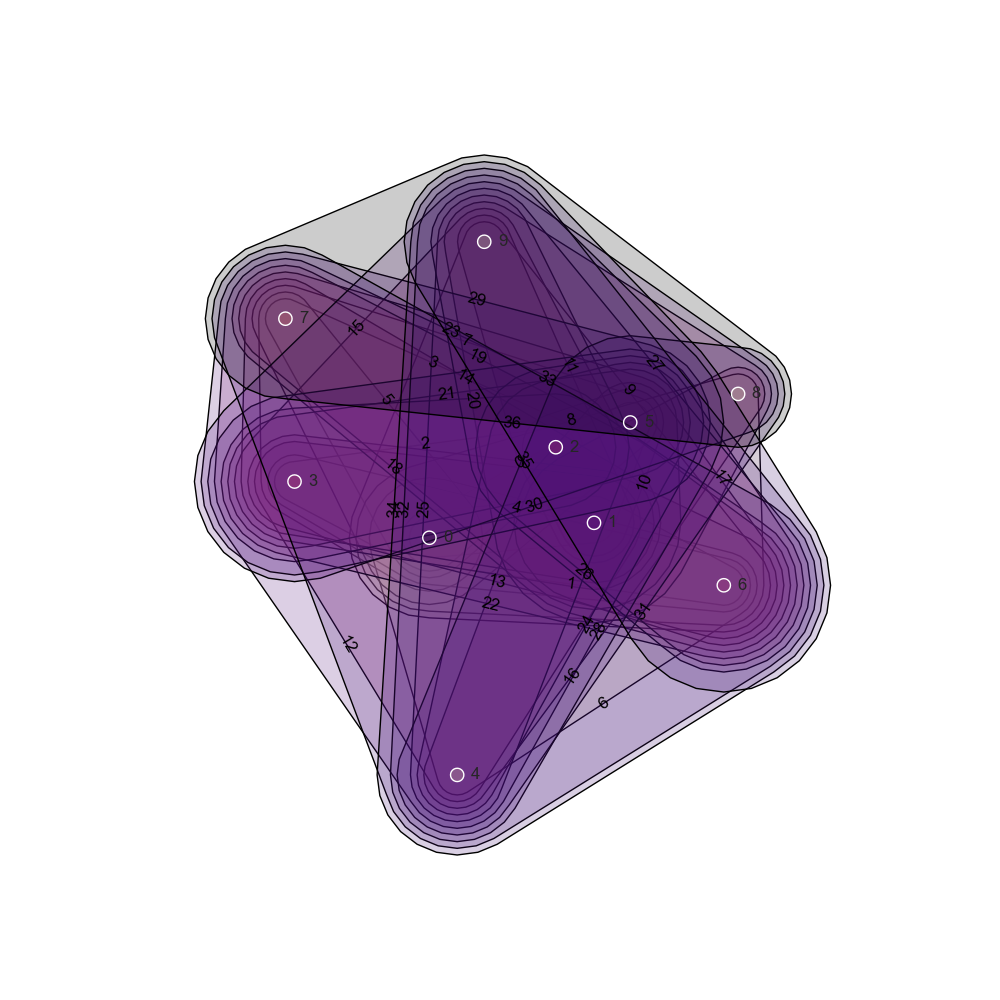}        \vspace{-\baselineskip}

    \caption{Generated Instances of a 2-layer model with $c = [1.5, 2.5], H = [0.25,1], \lambda = \log 3$. Left: $k = 2$, Right: $k = 3$.}
    \vspace{-\baselineskip}

    \label{fig:instances}
\end{figure}

The value of $\zeta = 2$ works well empirically, so we set $\zeta = 2$ from now on. We will refer to $c = ( c_i )_{i \in [L]}$ as the \emph{density parameters} (or \emph{core profile}) and to $H = (H_i)_{i \in [L]}$ as the \emph{breakpoints}. The models agree with the stochastic blockmodel of \cite{zhang2015identification}, namely for nodes that are closer to the core, their probability of jointly participating as a hyperedge is higher than a subset of nodes that are further from it. The density parameters $c$ give us a way to tweak the density between different levels of the graph, thus giving us flexibility to encode more complex structures with a constant overhead in terms of complexity, when the number of layers is constant (in our experiments $L \in \{ 1, 2 \}$). Fig.~\ref{fig:instances} shows instances generated from CIGAM. 

\section{Fast Algorithms} \label{sec:fast_algorithms}

Note that the na\"ive exact computation of the log-likelihood (LL), i.e. without taking into account the sufficient statistics of each hyperedge, requires flipping $\sum_{k = k_{\min}}^{k_{\max}} \binom n k$ coins in total which is highly prohibitive even for hypergraphs with very small $k_{\max}$ even if $n$ is moderate ($n > 100$)\footnote{In our experiments $k$ typically ranges between 2 and 25. Moreover in the worst case when $k \in [2, n]$ na\"ively computing the likelihood costs $O(2^n)$.}. In the same way, sampling hypergraphs from CIGAM also needs flipping $\sum_{k = k_{\min}}^{k_{\max}} \binom n k$ coins which makes sample generation inefficient as well. We also note that in the simple case of graphs with spatial features the work of \cite{jia2019random}, the authors approximate the likelihood and sampling of their model, where, in our case, we take advantage of the \emph{sufficient statistics} of each hyperedge to do \emph{exact inference in linear time}. Our approach follows the methodology used to sample Kronecker hypergraphs \cite{eikmeier2018hyperkron}, however the partitioning of the graph is \emph{significantly} different in both cases and requires careful calculation.  

More specifically, we observe that a graph generated by the CIGAM model can be broken down into multiple Erd\"os-R\'enyi graphs, whose block sizes and parameters we devise in Sec.~\ref{sec:hyperedge_partitioning}, and such partitions have efficient representations in terms of the LL as well as can be sampled by standard methods for sampling Erd\"os-R\'enyi graphs  \cite{ramani2019coin}.

\subsection{Hyperedge Set Partitioning} \label{sec:hyperedge_partitioning}

From the definition of the model, we deduce that the hyperedge set can be efficiently partitioned, both for sampling and inference, namely each edge in the multi-layer model is determined by $\max_{u \in e} r_u$ and $\min_{u \in e} r_u$. The edge set $E$ of a hypergraph that contains simplices of order $k \in [k_{\min}, k_{\max}]$ is partitioned as

\begin{equation*}
    E = \bigsqcup_{k \in [k_{\min}, k_{\max}], \; i \in [n], \; l \in [L]} E(k, i, l),
\end{equation*}

where $E(k, i, l) = \{ e \in E : |e| = k, i = \argmax_{u \in e} r_u, \phi(e) = l \}$. To sample the multi-layer model, we first assign a layer to each node $i$, assuming again that $r_1 \succ r_2 \succ \dots \succ r_n$, that is the layer that the current node belongs if its the node with the smallest rank in a given hyperedge. We call this matrix, which has increasing entries, $\texttt{layers[} n \texttt{]}$. We then form $N(i, l) = \{ i < j \le n | \texttt{layers[} j \texttt{]} = l  \} $. Now, a hyperedge $e$ needs two components: the \emph{largest} rank, which determines the exponent of $f(e)$ and the \emph{smallest} rank in $e$ that determines the layer that the hyperedge is in. We start by iterating on the ranks array and while fixing $i$ as the dominant node: we first select $j$ from $N(i, l)$ and then, we sample $k - 2$ nodes from $[i + 1, j - 1]$, for $j - i - 1 \ge k - 2$ for every $k \in [k_{\min}, k_{\max}]$. Therefore, a total of $\sum_{j \in N(i, l)} \binom {j - i - 1} {k - 2}$ $k$-simplices can belong to the partition where $i$ is the dominant rank node. We construct a $(k_{\max} - k_{\min} - 1) \times n \times L$ matrix which contains $|E(k, i, l)|$ for all $k \in [k_{\min}, k_{\max}]$, $i \in [n]$ and $l \in [L]$. We iterate over all $e \in E$ and increment the $(|e|, \argmax_{u \in e} \{ r_u \}, \texttt{layers[} \argmin_{u \in e} \{ r_u \} \texttt{]})$-th entry of the matrix. The complement sets have sizes

\begin{equation*}
    | \bar E(k, i, l) | = \sum_{j \in N(i, l)} \binom {j - i - 1} {k - 2} - |E(k, i, l)| \quad \forall k, i, l.
\end{equation*}

To simplify the summation, note that by the binomial identity $\binom {j - i - 1} {k - 2} = \binom {j - i} {k - 1} - \binom {j - i - 1} {k - 1}$, we calculate the sum $\sum_{j \in N(i, l)} \binom {j - i - 1} {k - 2}$ as 

\begin{equation} \label{eq:region_size}
    \sum_{j \in N(i, l)} \binom {j - i - 1} {k - 2} = \binom {j_{\max} (i, l) - i} {k - 1} - \binom {j_{\min} (i, l) - i - 1} {k - 1},
\end{equation}

with $j_{\max}(i, l) = \max_{j \in N(i, l)} j$, and $j_{\min} (i, l) = \min_{j \in N(i, l)} j$ and using the fact that $N(i, l)$ is a contiguous array. We can further validate that for the simple case that $k_{\min} = k_{\max} = k$, $L = 1$ we have that $j_{\min}(i, 1) = i + 1$ and $j_{\max} (i, 1) = n$ yielding $|\bar E(i, 1)| = \binom {n - i} {k - 1} - |E(i, 1)|$ as expected. As a generative model, each block requires throwing $E|(k, i, l)|$ balls where $|E(k, i, l)|$ follows a binomial r.v. with  $\binom {j_{\max} (i, l) - i} {k - 1} - \binom {j_{\min} (i, l) - i - 1} {k - 1}$ trials of bias $c_l^{-2+r_i}$ for a hyperedge of order $k$. We further simplify the number of trials by noting that $j_{\max}(i, l) = j_{\min}(i, l) + |N(i, l)| - 1$ since $N(i, l)$ is a contiguous array. We need $O(k_{\max}(m + n) + n \log n)$ to build the $|E(k, i, l)|$ and time $O(n (k_{\max} + L))$ to build $|\bar E(k, i, l)|$ giving a total preprocessing time of $T_{PRE} = O(n (k_{\max} + L) + k_{\max} m + n \log n)$. Usually (see Sec.~\ref{sec:experiments}), $k_{\max}$ and $L$ are constant which brings $T_{PRE}$ to $O(n \log n + m)$.  

\subsection{Exact Inference} \label{sec:inference}

Based on the hyperedge set partitioning of Sec.~\ref{sec:hyperedge_partitioning} we can compute the LL function $\log p(G, r | \lambda, c)$ as follows, assuming the elements of $r$ are \emph{sorted in decreasing order}:

\begin{equation}
\begin{split}
    \log p(G, r | \lambda, c ) = &  - \lambda \sum_{i \in [n]} r_i + n \log \lambda - n \log (1 - e^{-\lambda}) \\
    & + \sum_{i, l} \bigg [ \left ( \sum_{k} |E(k, i, l)| \right ) \log \left ( c_l^{-2+r_i} \right ) \\
    & + \left ( \sum_{k}|\bar E(k, i, l)| \right ) \log \left ( 1 - c_l^{-2+r_i} \right )  \bigg ].
\end{split}
\end{equation}

Then, the likelihood can be computed in $T_{LL} = O(nL)$ time by precomputing $\sum_{k} |E(k, i, l)|$ and $\sum_{k} |\bar E(k, i, l)|$. The total memory required is $O(nL)$ to store $\sum_{k} |E(k, i, l)|$ (resp. $\sum_{k} |\bar E(k, i, l)|$), $O(m)$ to store the edges, and $O(nk_{\max})$ to store the binomial coefficients. Therefore, the total memory cost is $O(n (k_{\max} + L) + m)$. Since the number of layers $L$ are constant compared to $n$, the time needed to compute the likelihood is $O(n)$.  

The breakpoints $ H_i $ are  \emph{hyperparameters} of the model. We impose constraints on the density of the edges, i.e. we want the graph to ``sparsify'' towards the periphery which can be encoded with the following domain $\mathcal K = \{ (\lambda, c) : 1 < c_1 < c_2 \dots < c_L \}$. We encode the domain $\mathcal K$ on the LL by considering the log-barrier function 
$\varphi \left (\lambda, c \right ) = \log (c_1 - 1) +  \sum_{i = 1}^{L - 1} \log (c_{i + 1} - c_i)$ 
which equals $- \infty $ for every $(\lambda, c) \notin \mathcal K$. Thus, we solve the following inference problem to get the optimal parameters of the model 

\begin{equation*}
    \max_{(\lambda, c) \in \mathcal K}  \log p (G, r | \lambda, c) \Leftrightarrow \max_{(\lambda, c) \in \mathbb R^{L + 1}} \log p(G, r | \lambda, c) + \varphi(\lambda, c).
\end{equation*}

\subsection{Learning Endogenous Ranks} \label{sec:learning_ranks}

When the rank vector $r$ is not provided we learn the endogenous (or \emph{latent}) ranks given a feature matrix $X \in \mathbb R^{n \times d}$ by using a learnable model $h ( \cdot | \theta )$ to get $r_i = h(x_i | \theta)$, which we replace on the LL as follows:

\begin{equation*}
\begin{split}
    \log p(G, X | \lambda, c, \theta ) & =  - \lambda \sum_{i \in [n]} h(x_i | \theta) + n \log \lambda - n \log (1 - e^{-\lambda}) \\
    & + \sum_{i, l} \bigg [ \left ( \sum_{k} |E(k, i, l)| \right ) \log \left ( c_l^{-2+h(x_i | \theta)} \right ) \\
    & + \left ( \sum_{k}|\bar E(k, i, l)| \right ) \log \left ( 1 - c_l^{-2+h(x_i | \theta))} \right )  \bigg ].
\end{split}
\end{equation*}

 In Sec.~\ref{sec:experiments}, we use a simple logistic model for determining the ranks, i.e. $r_i = h(x_i | w, b) = \sigma ( w^T x_i + b )$ where $\sigma (\cdot)$ is the sigmoid function. After training, we can use the learned parameters $\theta^*$ to extract embeddings that capture the \emph{endogenous ranks} of nodes in the graph. Besides, we can use centrality measures (degree centrality, [hypergraph] eigenvector centrality for $k$-uniform hypergraphs \cite{benson2019three}), PageRank etc.) to enrich the feature vectors. Here, the ranks $r_i$ need to be re-sorted after each forward call to $h$ which makes the total per-step cost for evaluating the LL equal to $T_{PRE} + T_{LL} + O(dn)$. 

\subsection{Choosing Hyperparameters} \label{sec:hyperparameters}

\begin{figure}[t]
    \centering
    \includegraphics[clip,trim={0.5in 0.5in 0.5in 0.5in}, width=0.49\columnwidth]{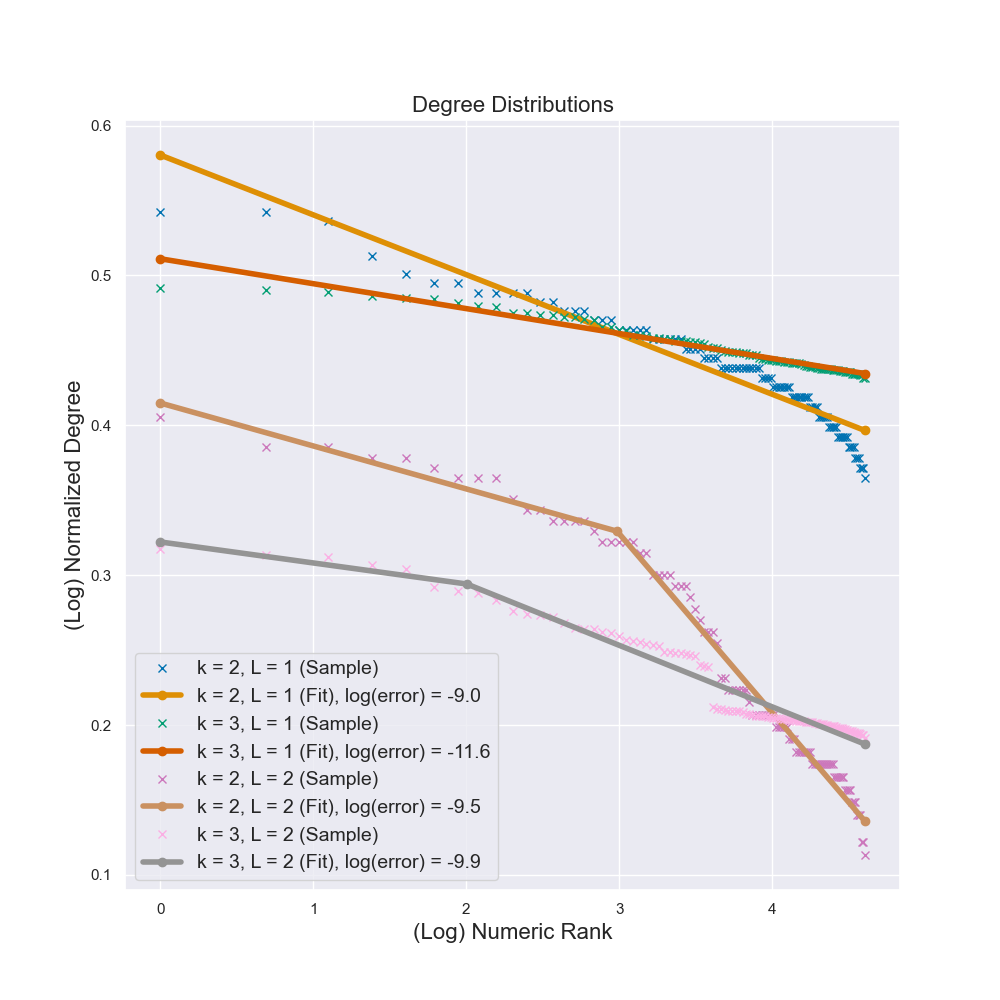}
    \includegraphics[clip,trim={0.5in 0.5in 0.5in 0.5in}, width=0.49\columnwidth]{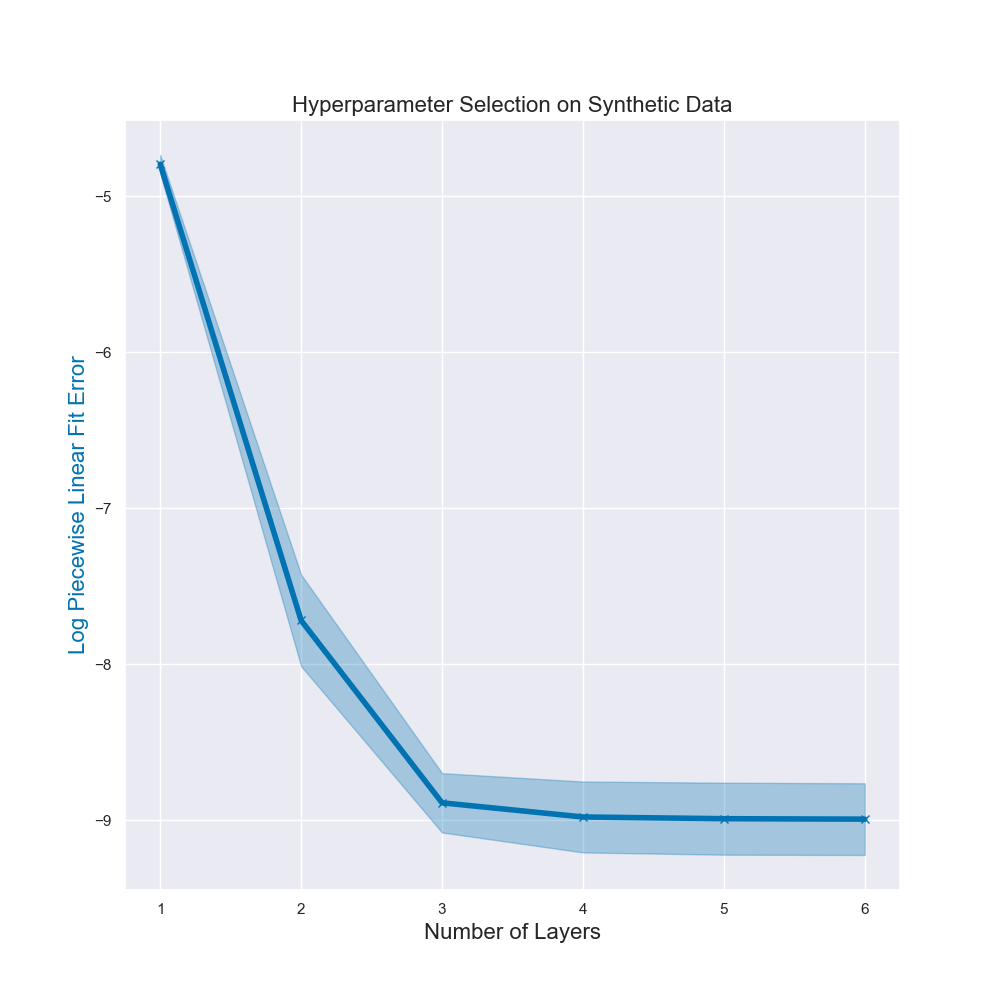}
    \vspace{-\baselineskip}
    \caption{(Left) Degree Plot for an instance with $k \in \{ 2, 3 \}, L \in \{ 1, 2 \}$ layer, $b = 3$, $H = 1$, $H$ split as powers of $1/2$ in $[0, 1]$ and $c$ split uniformly on $[1.5, 2.9]$ with p/w fits.
    (Right) Elbow plot for 10 3-layer simulated graphs.}
    \vspace{-\baselineskip}
   \label{fig:expected_degree_distributions}
\end{figure}

A question that arises when we fit a multi-layer CIGAM model is the following: \emph{How to choose the number of layers $L$ and the breakpoints $H$?} We observe that samples generated from CIGAM have roughly a piecewise linear form when the observed degree is plotted versus the degree ordering of a node in the degree ordering in a log-log plot (Fig.~\ref{fig:expected_degree_distributions}). This observation motivates the following heuristic for hyperparameter selection: given a hypergraph $G$ we calculate the degrees of all nodes, sort them in decreasing order and fit a piecewise linear function on the log-log scale. We then apply the \emph{elbow} criterion \cite{thorndike1953belongs} and choose the number of layers to be 

\begin{equation*}
    L_{\mathrm{pw}} = \underset {l \in [2, L_{\max} - 1]} {\mathrm{argmax}} \left \{ \frac {\log \pwerr^*(l) - \log \pwerr^*(l - 1)} {\log \pwerr^*(l + 1) - \log \pwerr^*(l)}  \right \}
\end{equation*}

where $\pwerr^*(l)$ is the minimum piecewise linear fit error when fitting a function that is a piecewise combination of $l$ lines. Roughly the rule says to pick the number of layers around which the ratio of the subsequent gradients is maximized. Then to identify the breakpoints we run grid search (i.e. likelihood ratio test/AIC\footnote{The difference between LR and the AIC between two models with layers $L$ and $L'$ is $L - L'$, so the two measures differ by at most 1, i.e. they are very close.}/BIC) on all feasible $L_{\mathrm{pw}}$ breakpoints.

\subsection{Exact Sampling} \label{sec:sampling}

Given the parameters $\lambda, c$ of a single-layer model, a reasonable question to ask is: How can we efficiently generate $k$-uniform -- and subsequently general hypergraphs -- samples from the model with parameters $\lambda, c$? Identically to computing the LL a nai\"ve coin flipping approach is computationally intractable; and can be exponential in the worst case. To mitigate this issue we use the \emph{ball-dropping} technique to generate edges as follows: 

\manualnumber{Step 1.} Generate the $n$ ranks $r \sim \mathrm{TruncatedExp}(\lambda, [0, 1])$ with inverse transform sampling in $O(n)$ time (assuming access to a uniform $[0, 1]$-variable in $O(1)$ time). Sort wrt. $\succ$.

\manualnumber{Step 2.} For each $i \in [n]$ in the order draw a random binomial variable $M_i \sim \mathrm{Bin} \left ( \binom{n - i} {k - 1}, c^{-2+r_i} \right )$. Drop $M_i$ balls, where each ball represents a hyperedge $e$ with $i \in e$ (see App.~\ref{app:sampling} for how hyperedges are sampled). 

We repeat the same process for various values of $k$ to create a graph with multiple orders. This technique runs much fastetr than $\binom n k$ (see \cite{ramani2019coin}). 
The same logic can be extended to the multi-layer model where instead of dropping $M_i \sim \mathrm{Bin} \big ( \binom {n - i} {k - 1},  c^{-2+r_i} \big )$ balls for each $i$ on the corresponding spots, we (more generally) throw $M_{i, l} \sim \mathrm{Bin} \left ( \binom {j_{\max}(i, l) - i} {k - 1} - \binom {j_{\min} (i, l) - i - 1} {k - 1}, c_l^{-2+r_i} \right )$ for every $i \in [n]$ and $l \in [L]$ where each ball spot is chosen by throwing the first ball between $j_{\min} (i, l)$ and $j_{\max} (i, l)$ and the rest $k - 2$ balls between $i$ and $j_{\max} (i, l)$. We again use a rejection sampling mechanism to sample from this space. Finally, to sample a hypergraph with a simplex order range between $k_{\min}$ and $k_{\max}$ we repeat the above process for every $k \in [k_{\min}, k_{\max}]$ and take the union of the produced edge sets. Our implementation contains ball-dropping techniques for the general case of multiple layers, however here we present the single layer case for clarity of exposition.

\section{``Small-core'' Property} \label{sec:small_core}

The model we are using serves as a generalization of hierarchical random graph models to random hypergraphs. In those models, see e.g. \cite{papachristou2021sublinear}, it is often a straightforward calculation to bound the size of the core. Nevertheless, trivially carrying out the same analysis on hypergraphs does not work since the combinatorial structure of the problem changes cardinally, which highlights the needs of new analysis tools in order to be proven. 

In detail, in order to characterize the properties of the core of networks generated with CIGAM, we ask the following question: Given a randomly generated $k$-uniform hypergraph $G$ generated by CIGAM, \emph{what is the size of its core}? In our regime the core is a subset $C$ of the vertices with the following properties:

\manualnumber{1.} Nodes within the core set $C$ are ``tightly'' connected with respect to the rest of the graph. 

\manualnumber{2.} Nodes within the core set $C$ form a dominating set for $G$ \emph{with high probability} (i.e. w.p. $1 - O(1/n)$).

For the former requirement, it is easy to observe that the induced subhypergraph that contains nodes with $r_i \ge t$ for some appropriately chosen $t$ would form the most densely connected set wrt. the rest of the network. For the latter requirement, we use a probabilistic argument to characterize the core. 

Clearly, members of the core are responsible for covering the periphery of the graph, i.e. each peripheral node has at least one hyperedge in the core. Thus the size of the core is the number of nodes that are needed to cover the periphery with high probability. We analyze the core size of the multi-layer model by constructing coupling with a single layer model that generates $k$-uniform hypergraphs with density $c_L$ (that corresponds to the ``sparsest'' density). Then, we start with a threshold $t \in [0, 1]$ and we let $N_k(t)$ be the number of $(k - 1)$-combinations that have at least one rank value $\ge t$. Then, we need to determine the value of $t$ such that \emph{(i)} $N_k(t)$ exceeds some quantity $N_0$ with high probability; \emph{(ii)} the nodes with $r_j \ge t$ form an almost dominating set (serving as the core of the graph) with probability $1 - O(1 / n)$ conditioned on $N_k(t) \ge N_0$. Proving \emph{(ii)} proceeds by calculating the (random) probability that a node is not dominated by a \emph{core hyperedge}, i.e. a hyperedge that has at least one node with $r_j \ge t$, and then taking a union bound over the nodes and setting the resulting probability to be at most $1 / n$, yielding the desired lower bound $N_0$. Now, given that we know $N_0$ we want to set it in a value such that $N_k(t) \ge N_0$ with probability at least $1 - 1 / n$. Observing that $N_k(t)$ obeys a combinatorial identity involving $N_2(t)$ which is a binomial r.v. and by using the Chernoff bound on $N_2(t)$ we can get a high probability guarantee for $N_k(t) \ge N_0$ by setting $N_0$ appropriately.  
Finally, we set the threshold $t$ such that the probability of having a core at threshold $t$ is at least $1 - O(1/n)$. We present the Theorem (proof in App.~\ref{app:core})

\begin{theorem}[Core Size] \label{theorem:core_size}
    Let $G$ be a $k$-uniform hypergraph on $n \ge 2$ nodes with $k < n$ generated by a CIGAM model with $L$ layers and parameters $1 < c_1 \le c_2 \dots \le c_L < e^{\lambda}$ for $\lambda < \tfrac {\ln (n / 72)} 4$. Then, with probability at least $1 - 2 / n$ the graph has a core at a threshold $t$ such that $\tfrac {2 \log n} {c_L^{-2+t}} = \binom {n} {k - 1} - \binom {n F(t) + \sqrt {n \log n / 2}} {k - 1}$ with size $\til O(\sqrt n)$. $\binom x y$ is the generalized binomial coefficient. 
\end{theorem}

\begin{figure}
    \centering
    \includegraphics[width=0.85\columnwidth]{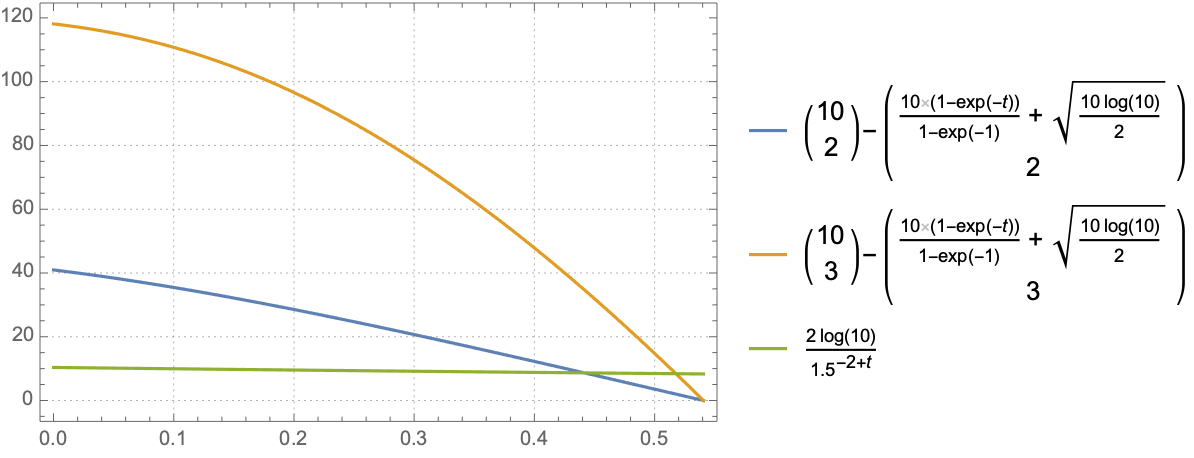}
        \vspace{-\baselineskip}
    \caption{Core threshold functions from Thm. \ref{theorem:core_size} for $c_L = 1.5, \lambda = 1, k \in \{ 3, 4 \}$, and $t \in [0, F^{-1} (1 - \sqrt {\log n / (2n)}  ) ] $ (x-axis).}
    \label{fig:core_size}
\end{figure}

Fig.~\ref{fig:core_size} depicts the (theoretical) threshold $t$ for a 3-uniform and a 4-uniform hypergraph on 10 nodes. In App. \ref{app:core} we also plot the empirical thresholds of CIGAM-generated hypergraphs with $n = 200$. As $k$ increases the threshold $t$ moves to the right and therefore the core becomes smaller. 

\section{Experiments} \label{sec:experiments}

We first validate our model's ability to recover the correct parameters on synthetically generated data, as well as the efficiency of the proposed sampling method. We then perform experiments on small-scale graphs and show that the recovered latent ranks (and their subsequent ordering) can accurately represent the degree structure of the network. Finally, we do experiments with large-scale hypergraph data where we evaluate and compare our model with (generalized) baselines with respect to their abilities to fit the data.  

We implement \emph{Point Estimation (MLE/MAP)} and \emph{Bayesian Inference (BI)} algorithms as part of the evaluation process, which are available in the code supplement. App. \ref{app:implementation} describes the specifics of each implementation and the design choices, and Tab.~\ref{tab:complexity} shows the costs of fitting CIGAM on various occasions. 

    
    

\begin{table}
    \centering
    \footnotesize
    \caption{Time complexity of fitting CIGAM for preprocessing (PP) and log-likelihood (LL). Here $r_i = \sigma (w^T x_i + b)$.}        \vspace{-\baselineskip}
    \begin{tabular}{lllll}
        \toprule
        Ranks & Method & \# Params & PP & LL \\
        \midrule 
        Known & All & $L + 1$ & $T_{PRE}$ & $T_{LL}$ \\
        Exogenous & BI & $L + 1 + n$ & -- & $T_{PRE} + T_{LL}$ \\
        Endogenous & MLE, MAP &  $L + d + 2$ & -- & $T_{PRE} + T_{LL} + O(dn)$ \\
        \bottomrule
    \end{tabular}
    \vspace{-\baselineskip}
    \label{tab:complexity}
\end{table}

\subsection{Experiments on Synthetic Data}

\begin{figure}
    \centering
    \includegraphics[clip,trim={0.6in 0.6in 0.6in 0.6in}, width=0.9\columnwidth]{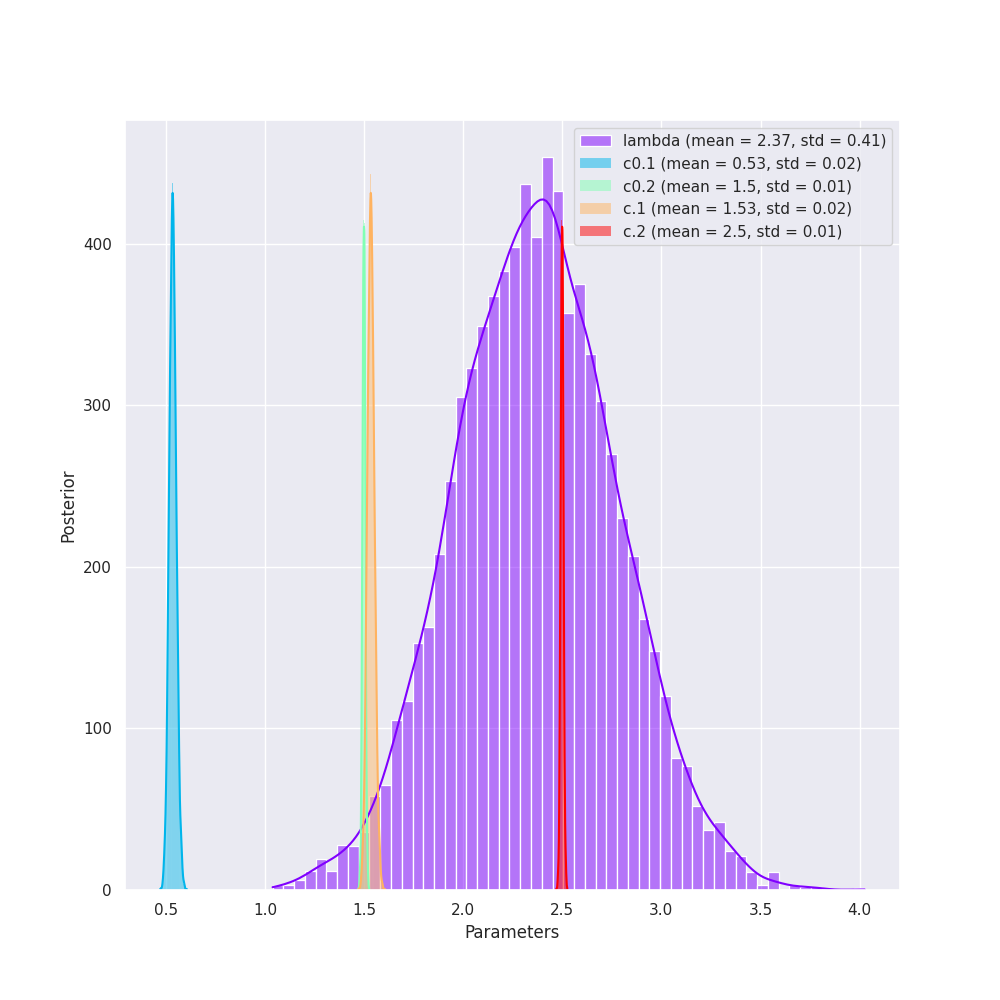}
    \includegraphics[clip,trim={0.6in 0.6in 0.6in 0.6in}, width=0.9\columnwidth]{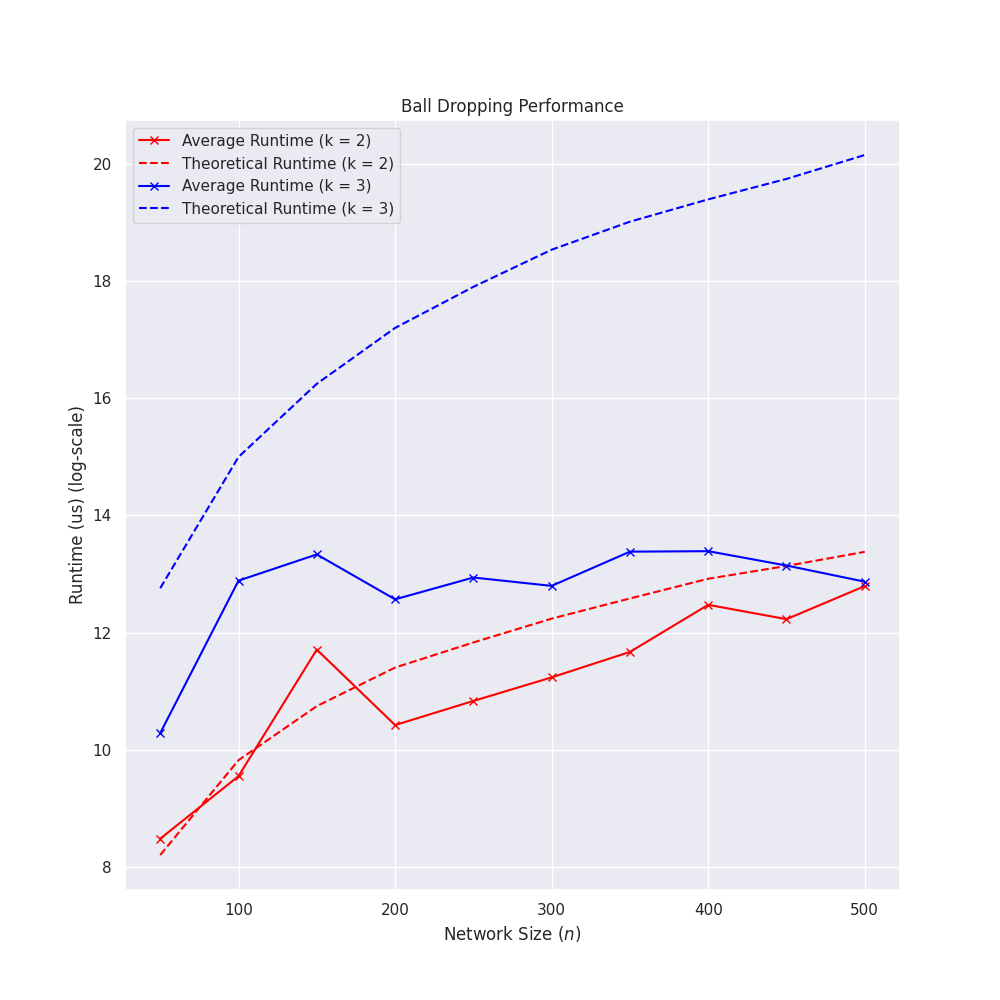}
    \caption{(Top) Parameter recovery for $n = 100, k \in \{ 2, 3 \}, c = [1.5, 2.5], \lambda = 2.5$. Legend: \texttt{c0} corresponds to off-by-one parameters and \texttt{c} corresponds to the actual parameters. (Bottom) Average Runtime of Sampling Using the Ball Dropping Method for hypergraphs of orders $k \in \{ 2, 3 \}$ and $50-500$ nodes with a step of 50 nodes for an 1-layer instance with $b = 3, c = 1.5$. The dashed line is the function is the expected number of edges of a $k$-uniform CIGAM hypergraph.}
    \vspace{-\baselineskip}

    \label{fig:synthetic_data}
\end{figure}

\noindent \textbf{Sampling.} Fig.~\ref{fig:synthetic_data} shows the performance of the ball-dropping method on 2-order and 3-order hypergraphs for graphs with 50-500 nodes with a step of 50 nodes where for each step we sample 10 graphs and present the aggregate statistics (mean and 1 standard deviation).

\noindent \textbf{Inference.} In Fig.~\ref{fig:synthetic_data}, we generate a 2-Layer graph with $n = 100$ nodes, $k \in \{ 2, 3 \}$, $\lambda = 2.5$, $c = [1.5, 2.5]$, $H = [0.5, 1]$, and use non-informative priors for recovery. Our algorithm can successfully recover the synthetic data.

\subsection{Recovering the Degree Structure of small-scale graphs}

\begin{figure}
    \centering
    \subfigure[world-trade \cite{de2018exploratory} ($n = 76, m = 845$)]{\includegraphics[width=0.49\columnwidth]{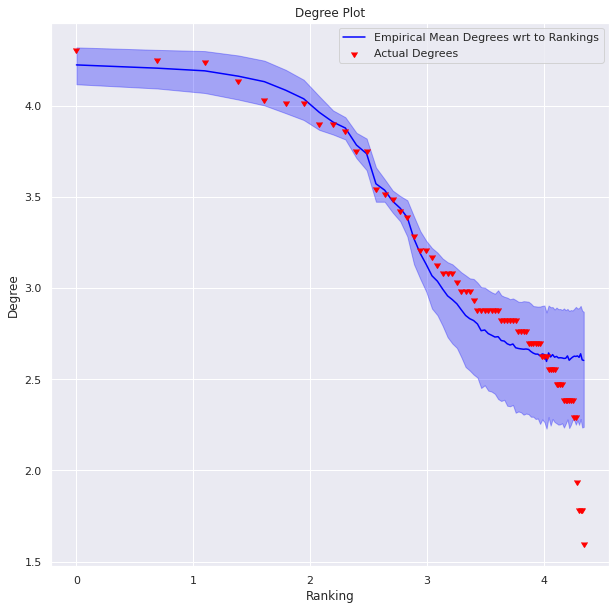}}
    \subfigure[c-elegans \cite{Kaiser-2006-placement} ($n \! = \! 279, m \! = \! 1.9 \mathrm K)$]{\includegraphics[width=0.49\columnwidth]{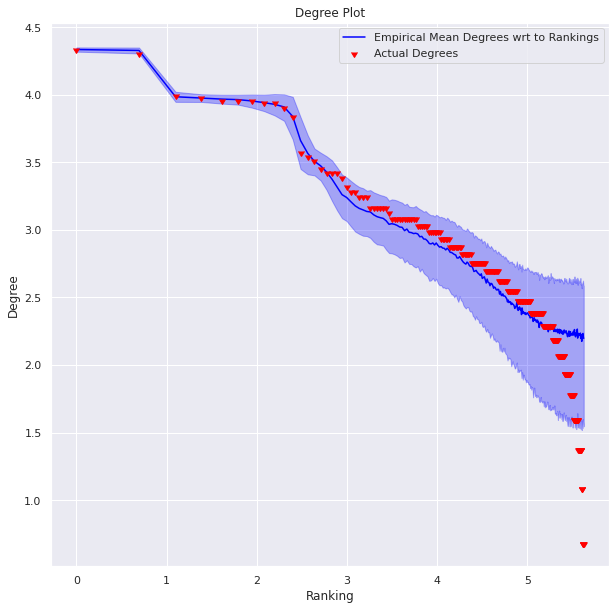}}
    \subfigure[history \cite{clauset2015systematic} ($n \! = \! 145, m \!  = \! 2 \mathrm K$)]{\includegraphics[width=0.49\columnwidth]{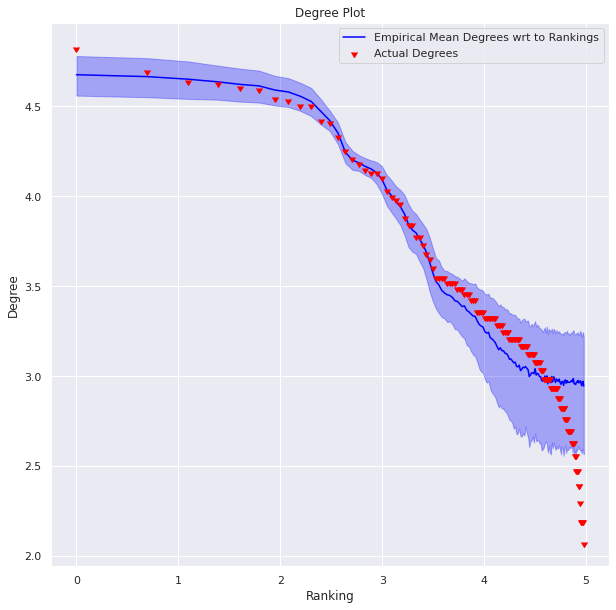}}
    \subfigure[business \cite{clauset2015systematic} ($n \! = \! 113, m \! = \! 3 \mathrm K)$]{\includegraphics[width=0.49\columnwidth]{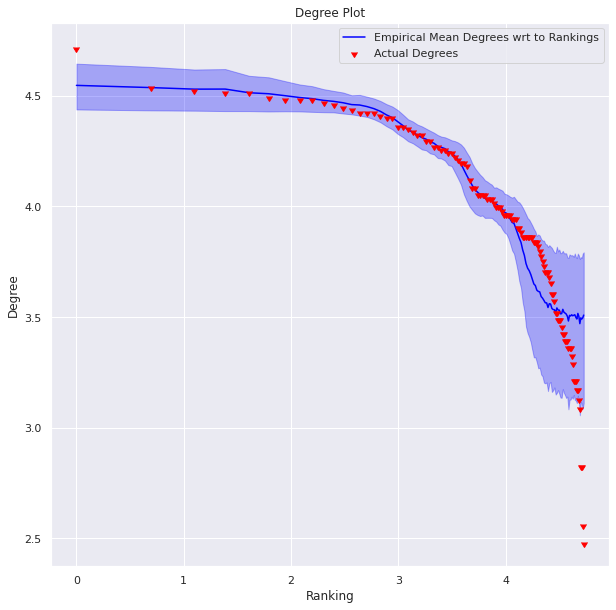}}
    \caption{Recovery of the Degree Structure for world-trade, c-elegans, history-faculty, and business-faculty datasets.}
    \label{fig:degree_structure}
\end{figure}

We perform BI on world-trade, c-elegans, history-faculty, and business-faculty using $L = 1$ layer, a $\mathrm{Gamma}(2, 2)$ prior for $\lambda$, and a $\mathrm{Pareto}(1, 2)$ prior for $c$. In Fig.~\ref{fig:degree_structure}, we order the actual degrees of the graphs in decreasing order and for every draw of the vector $r$ from the posterior (using MCMC with $N = 1 \text{K}$ samples) as follows: (i) We sort the entries of $r$ in decreasing order. Let $\{ \pi \}_{i \in [N]}$ be the corresponding permutations of the nodes. For each $i \in [n]$ we calculate the mean and the standard deviation of $\begin{pmatrix} \texttt{degrees[} \pi_{1, i} \texttt{]}, \dots, \texttt{degrees[} \pi_{N, i} \texttt{]} \end{pmatrix}^T$\footnote{I.e. the degree of the node which is first in the $i$-th ranking is added to the 1st position of the x-axis etc.}. The scales of Fig.~\ref{fig:degree_structure} are log-log. We observe that the degrees as they are determined by the ranks are consistent with the actual degree sequence. This suggests that core-periphery organization agrees with the degree centralities, as in \cite{papachristou2021sublinear}. 

\subsection{Experiments with large-scale data}

\noindent \textbf{Datasets.} We perform experiments with publicly available datasets. 

\begin{table}[]
    \footnotesize
    \centering
    \caption{Dataset Statistics. $n_{LCC}, m_{LCC}$ refer to the size of the LCCs, $n_{dt}, m_{dt}$ refers to the size of the LCCs after removing nodes with degree $\le 4$, $n_{kc}, m_{kc}$ refers to the size of the 2-core of the LCC.}
    \label{tab:dataset_statistics}
    \vspace{-\baselineskip}
    \begin{tabular}{lrrrrrrr}
    \toprule
    Dataset & $k_{\max}$ & $n_{LCC}$ & $m_{LCC}$ & $n_{dt}$ & $m_{dt}$ & $n_{kc}$ & $m_{kc}$  \\
    \midrule
    c-MAG-KDD & 22 & 2.7K & 1.4K & 47 & 40 & -- & -- \\
    t-ask-ubuntu & 6 & 6.2K & 7.8K & 664 & 1.7K & 50 & 33 \\
    t-math-sx & 7 & 26.4K & 66.1K & 5.4K & 39.2K & 100 & 58 \\
    t-stack-overflow & 10 & 164.9K & 306.4K & 34.3K & 140.4K & 7.0K & 6.0K \\
    ghtorrent-p & 7 & 78.4K & 74.8K & 10.8K & 14.8K & 3.4K & 2.7K \\
    \bottomrule
    \end{tabular}
    \vspace{-\baselineskip}
\end{table}

\begin{table*}[]
    \centering
    \footnotesize
    \caption{Hypergraph Experiments. We use SGD with step-size 0.001 for CIGAM, and SGD with stepsize of 1e-6 for Logistic-CP, for 10 epochs. For HyperNSM we use $a = 10$, $p = 20$, and $\xi(e) = 1 / |e|$. For evaluating the log-likelihood of Logistic-CP and HyperNSM we use $|\mathcal B| = 0.2 \bar m$ negative samples. Best likelihood is in bold. \dag{} = LL cannot be computed.}        \vspace{-\baselineskip}
    \begin{tabular}{l|rrrrr|rrrrr}
        \toprule
        Dataset & CIGAM & $c^*$ & $\lambda^*$ & Logistic-CP & HyperNSM  & CIGAM & $c^*$ & $\lambda^*$ & Logistic-CP & HyperNSM  \\
        \midrule
        & \multicolumn{5}{c}{Degree Threshold + LCC} & \multicolumn{5}{c}{LCC + 2-core} \\
        \midrule 
         & \multicolumn{10}{c}{Exogenous ranks} \\
        \midrule 
        coauth-MAG-KDD & \bf{-1334} & [3.0e+4] & 1.4 & -2025 & -2.5e+6 &  -- & -- & -- & -- & -- \\
        threads-ask-ubuntu & \bf{-1.3e+5} & [9.5e+9] & 3.5 & \dag & \dag & \bf{-166} & [11.94] & 1.5 & -1510 & -1073   \\
        threads-math-sx & \bf{-4.2e+6} & [2.3e+20] & 8.7 & \dag & \dag & -1736 & [5.2, 31.1] & 1.8 & \bf{-1478} & -1.4e+5 \\
        threads-stack-overflow & \bf{-2.0e+7} & [1.8e+20, 8.3e+27] & 4.8 & \dag & \dag &  \bf{-1.2e+5} & [2.1e+5] & 5.1 & -1.0+e7 & \dag \\
        ghtorrent-projects & \bf{-1.2e+6} & [7.8e+15, 5.9e+20] & 4.4 & \dag & \dag & \bf{-2.6e+5} & [9.4e+18] & 3.6 & \dag & \dag \\ 
        \midrule
        & \multicolumn{10}{c}{Endogenous (Learnable) ranks} \\
        \midrule
        coauth-MAG-KDD & -1.8e+5 & [32.4] & 1.1 & \bf{-2024} & -2.5e+6 & -- & -- & -- & -- & -- \\
        threads-ask-ubuntu & \bf{-1.6e+5} & [9.6e+6] & 1.1 & \dag & \dag & -2079 & [1.1] & 1.1 & \bf{-1659} & -1075 \\
        threads-math-sx & -4.2e+6 & [2.3e+20] & 11.3 & \dag & \dag & \bf{-3.8e+4} & [2.8] & 1.1 & -5.8e+4 & -1.4e+5 \\
        threads-stack-overflow & \bf{-1.9e+7} & [4.1e+29] & 26.8 & \dag & \dag & \bf{-4.0e+5} & [8.6e+8] & 1.1 & \dag & \dag \\
        ghtorrent-projects & \bf{-1.2e+6} & [2.9e+22] & 2.8 & \dag & \dag & \bf{-2.1e+5} & [9.4e+20] & 1.1 & \dag & \dag \\
        \bottomrule
    \end{tabular}
    \vspace{-\baselineskip}
    \label{tab:hypergraph_experiments}
\end{table*}

\manualnumber{1.} \emph{coauth-MAG-KDD.} Contains all papers published at the KDD conference and are included in the \emph{Microsoft Academic Graph v2}~\cite{tang2008arnetminer, sinha2015overview}. We also include data for each of the authors' number of citations, h-index, number of publications and use the R package \texttt{amelia} \cite{honaker2011amelia} to impute missing data at rows where at least one of the columns exists after applying a log-transformation. 
    
\manualnumber{2.} \emph{ghtorrent-projects.} We mined timestamped data from the \texttt{ghtorrent} project \cite{gousios2012ghtorrent, Gousi13} and created the ghtorrent-projects dataset where each hyperedge corresponds to the users that have push access to a repository. We used features regarding: number of followers on GitHub, number of commits, number of issues opened, number of repositories created, and the number of organizations the user participates at. 
    
\manualnumber{3.} \emph{threads-\{ask-ubuntu, math-sx, stack-overflow\} \cite{benson2018simplicial}.} Nodes are users on \texttt{askubuntu.com}, \texttt{math.stackexchange.com}, and \texttt{stackover\-flow.com}. A simplex describes users participating in a thread that lasts $\le 24$ hours. We observe that there is a high concentration of (non-engaged) users with reputation 1 and then the next peak in the reputation distrubution is at reputation 101. This bimodality is explained since Stack Exchange gives users an engagement bonus of 100 for staying on the platform. Therefore, we filter out all the threads at which non-engaged users (i.e. users with reputation less than 101) participate. We keep the (platform-given) reputation (as the ranks), the up-votes, and the down-votes of each user as her features. 

\noindent \textbf{Deduplication.} We keep each appearing hyperedge exactly~once. 

\noindent \textbf{Outlier Removal \& Feature Pre-processing.}  We filter \emph{outliers} from the data in two ways. This is done in order to guarantee the numerical stability of the fitting algorithms. In the former filtering (Degree Threshold + LCC), we remove all nodes with degree $< 4$ and then find the Largest Connected Component (LCC) of the resulting data. In the latter filtering (LCC + 2-core) we first find the LCC of the hypergraph and then keep the 2-core\footnote{The $\delta$-core of $G$ is a subgraph $G'$ such that all nodes in $G'$ have degree $\ge \delta$.} within it. The statistics of the post-processed datasets can be found at Tab.~\ref{tab:dataset_statistics}.

In the experiments considering exogenous ranks only, we take logarithms (plus one) of the exogenous ranks and min-max normalize the results to lie in $[0, 1]$. For the learning task (i.e. endogenous ranks), we perform standard $Z$-normalization (i.e. subtract column means and divide by the column stds.) in lieu of min-max normalization.

\begin{table*}[t]
    \centering
    \footnotesize
    \caption{Projected Graph Experiments. We use SGD with step-size 0.001 for CIGAM, and SGD with stepsize of 1e-6 for Logistic-CP, for 10 epochs. For evaluating the log-likelihood of Logistic-CP, and Logistic-TH $(\alpha = 10, p = 20)$ we use $|\mathcal B| = 0.2 \bar m$ negative samples. Best likelihood in bold. \dag{} = LL cannot be computed.}
    \vspace{-\baselineskip}
    \begin{tabular}{l|rrrrr|rrrrr}
        \toprule
        Dataset & CIGAM & $c^*$ & $\lambda^*$ & Logistic-CP & Logistic-TH &  CIGAM & $c^*$ & $\lambda^*$ & Logistic-CP & Logistic-TH \\
        \midrule
        & \multicolumn{5}{c|}{Degree Threshold + LCC} & \multicolumn{5}{c}{LCC + 2-core} \\
        \midrule 
        & \multicolumn{10}{c}{Exogenous ranks} \\
        \midrule 
        coauth-MAG-KDD & \bf{-532} & [12.2] & 1.4 & -1532 & -1419 & -- & -- & -- & -- & -- \\
        threads-ask-ubuntu & -2.8e+4 & [1.9e+3] & 3.5 & \bf{-1.3e+4} & -2.4e+5 & \bf{-166} & [11.94] & 1.5 & -1510 & -1329 \\
        threads-math-sx & \bf{-8.5e+5} & [116.1, 2.0e+3] & 8.7 & -6.1e+6 & -1.6e+7 & \bf{-564} &  [2.7, 4.9] & 1.8 & -1376 & -5471 \\
        threads-stack-overflow & \bf{-4.0e+6} & [5.1e+6] & 4.8 & \dag & \dag & \bf{-1.2e+5} & [2.1e+5] & 5.1 & \dag & -2.7e+7 \\
        ghrtorrent-projects & \bf{-4.9e+6} & [1.7e+6] & 4.4 & \dag & \dag & \bf{-7.4e+5} & [359.3, 951.3] & 3.6 & \dag & -1.8e+6 \\  
        \midrule
        & \multicolumn{10}{c}{Endogenous (Learnable) ranks} \\
        \midrule
        coauth-MAG-KDD & -2171 & [1.1] & 1.1 & -1680 & \bf{-1458} & -- & -- & -- & -- & -- \\
        threads-ask-ubuntu & -5.0e+4 & [3.2] & 1.1 & -3.4e+4 & -2.8e+5 & \bf{-185} & [12.1] & 1.3 & -1666 & -1329 \\
        threads-math-sx & -1.3e+6 & [1.2e+5] & 10.7 & \dag & \dag & \bf{-402} & [45.2] & 1.7 & -1926 & -5471 \\
        threads-stack-overflow & \bf{-3.8e+6} & [5.2e+6] & 20.7 & \dag & \dag & \bf{-1.2e+5} & [2.1e+5] & 1.9 & -1.0e+7 & -2.7e+7 \\
        ghtorrent-projects & \bf{-7.1e+5} & [5.1e+5] & 4.4 & \dag & \dag & \bf{-5.7e+4} & [1.4e+4] & 4.5 & -1.9e+5 & -1.8e+6 \\
        \bottomrule
    \end{tabular}
    \label{tab:graph_experiments}
\end{table*}

\begin{figure*}
    \centering
    \includegraphics[clip,trim={0.5in 0.5in 0.6in 0.6in}, width=0.49\columnwidth]{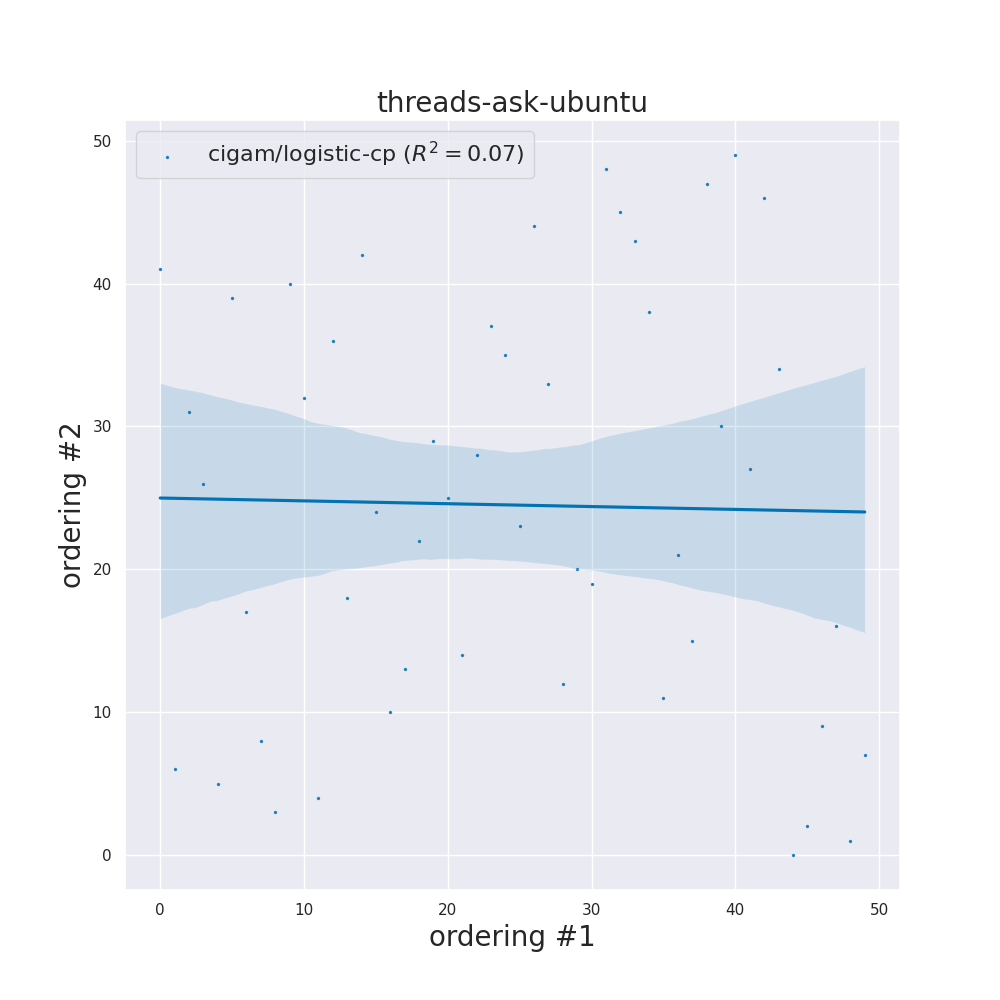}
    \includegraphics[clip,trim={0.5in 0.5in 0.6in 0.6in}, width=0.49\columnwidth]{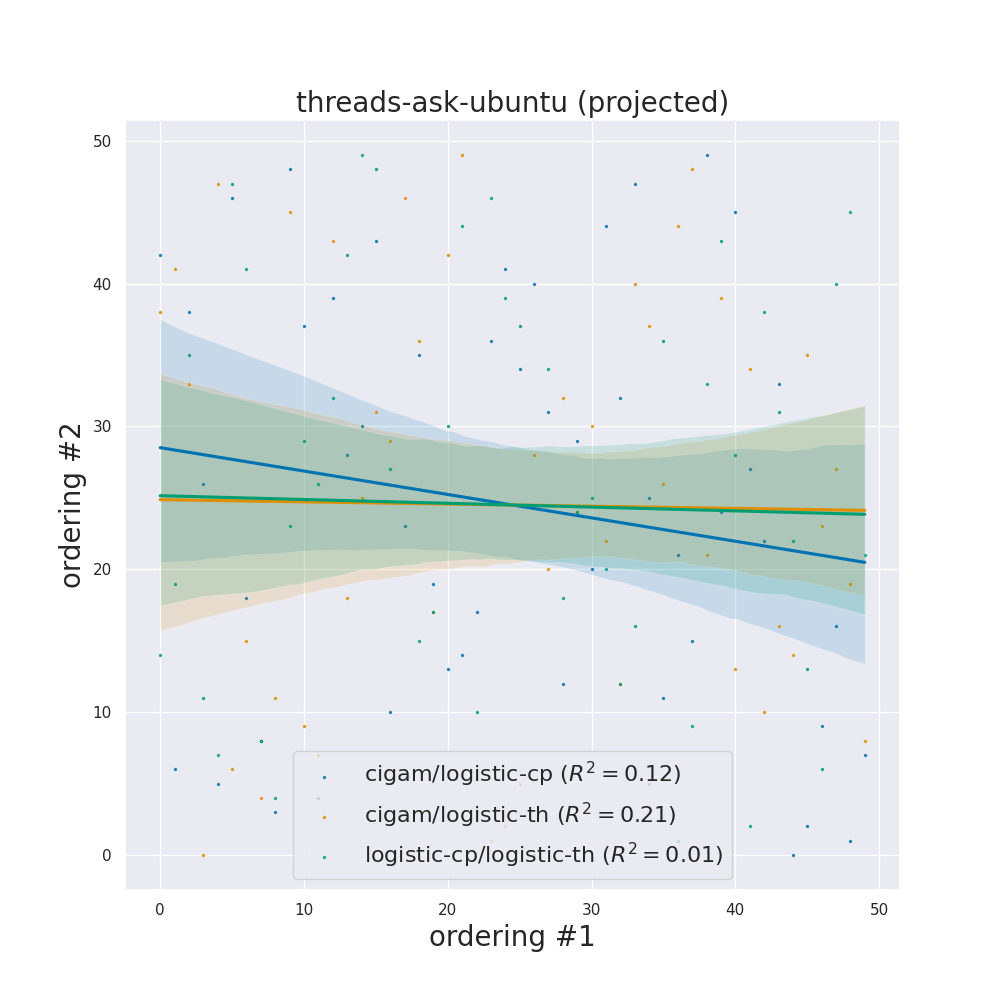}
    \includegraphics[clip,trim={0.5in 0.5in 0.6in 0.6in}, width=0.49\columnwidth]{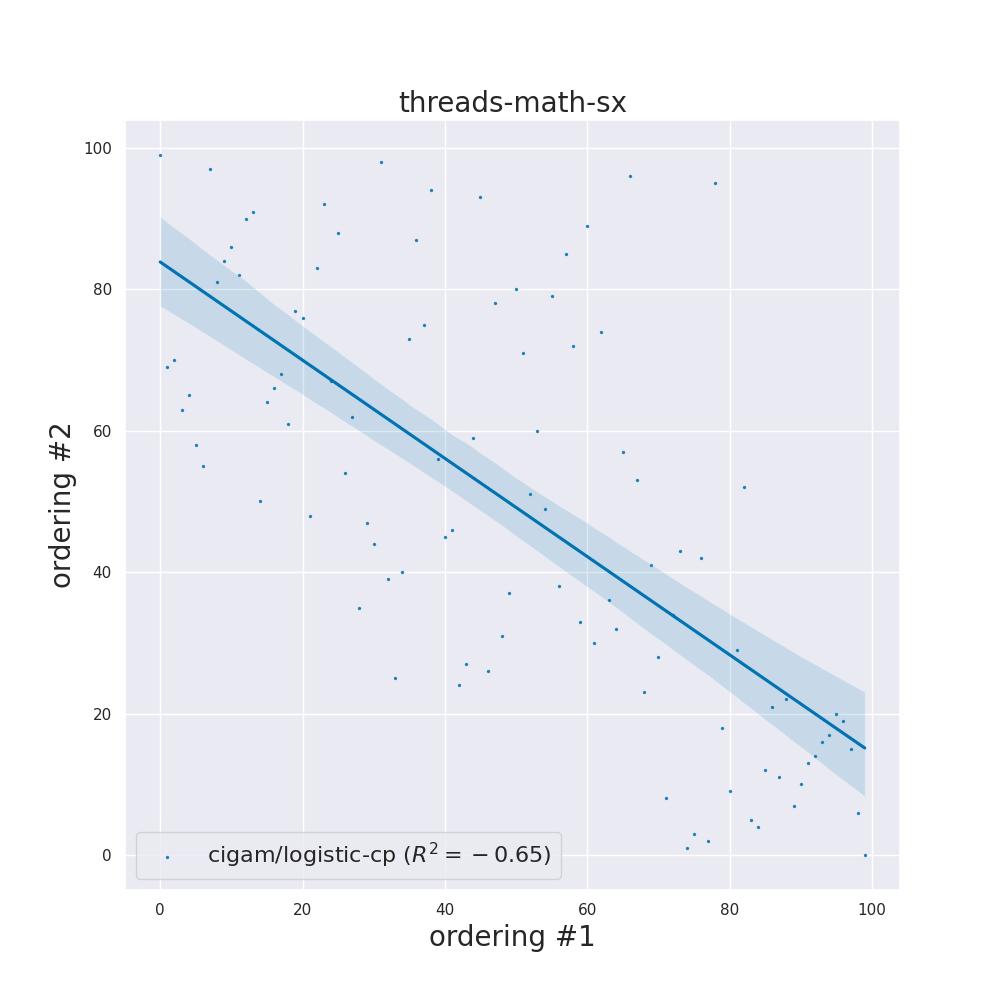}
    \includegraphics[clip,trim={0.5in 0.5in 0.6in 0.6in}, width=0.49\columnwidth]{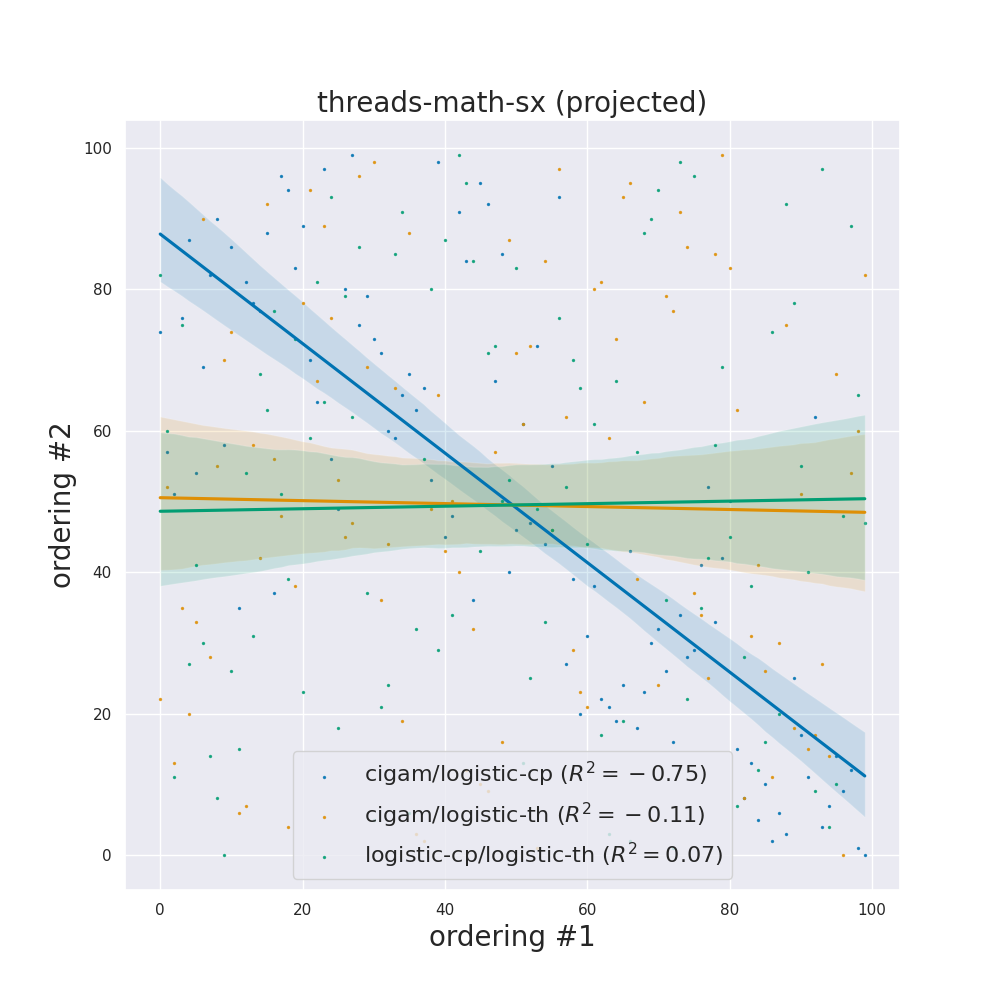}
    \vspace{-\baselineskip}
    \caption{Ranking comparison between algorithms (Tabs.~\ref{tab:hypergraph_experiments}, \ref{tab:graph_experiments}) for threads-\{ask-ubuntu, math-sx\} datasets (LCC + 2-core).}
    \vspace{-\baselineskip}
    \label{fig:spearman_algorithms}
\end{figure*}

\noindent \textbf{Hypergraph Experiments.} We compare CIGAM with the \emph{baseline} (without spatial dependencies) logistic model of \cite{jia2019random}. This model is also known as the $\beta$-model \cite{stasi2014beta,wahlstrom2016beta} and has been already generalized to hypergraphs (see \cite{stasi2014beta}). However, the inference algorithm of \cite{stasi2014beta} suffers from combinatorial explosion since for every node, the fixed-point equations require a summation over $\binom {n - 1} {k - 1}$ terms for a $k$-uniform hypergraph which makes the inference task infeasible for the datasets we study. 

Logistic-CP generates independent hyperedges based on the sum of core scores $\sum_{i \in e} z_i$ for a hyperedge $e$, i.e. $e$ is generated with probability $\rho(e) = \sigma \left ( \sum_{i \in e} z_i \right )$ where $\sigma(z) = \tfrac {1} {1 + e^{-z}}$. That corresponds to a LL $\log p (G | z) = \sum_{e \in E} \log \rho(e) + \sum_{\bar e \notin E} \log (1 - \rho(e))$. In contrast to our model, computing the likelihood of Logistic-CP and its gradient exactly is -- contrary to the case of CIGAM -- intractable since it requires exhaustively summing over $\bar e \notin E$ which can be very large. To approximate the log-likelihood we use negative sampling by selecting a batch $\mathcal B$ as follows: We sample a hyperedge order $K \in [k_{\min}, k_{\max}]$ with probability $\Pr [K = k] = \tfrac {\bar m_k} {\bar m}$ where $\bar m_k = \binom n k - m_k$ denotes the number of negative hyperedges of order $k$ and then given the sampled order $K$ we sample a negative hyperedge $\bar e$ uniformly from the set of non-edges of order $k$. That uniform sampling scheme has a probability of selecting a hyperedge $\bar e$ equal to $\tfrac {|\mathcal B|} {\bar m}$ (App.~\ref{app:sampling} describes the sampling algorithm). We use the following unbiased estimator of the LL based on this sampling scheme, 

\begin{equation}
    \hat \ell (G | z) = \sum_{e \in E} \log \rho(e) + \frac {\bar m} {|\mathcal B|} \sum_{\bar e \sim \mathcal B} \log (1 - \rho(\bar e)),
\end{equation}

where its easy to confirm that $\ev {} {\hat \ell} = \log p(G|\theta)$, and that $\var {} {\hat \ell} = \frac {\bar m - |\mathcal B|} {|\mathcal B|} \sum_{\bar e \in \bar E} \log^2(1 - \rho(\bar e))$, which is at most $\tfrac {(\bar m - | \mathcal B|) \bar m} {| \mathcal B |}$ for $ \rho(\bar e) \in [0, 0.51]$ (true for real-world datasets). Letting $| \mathcal B | = \alpha \bar m$ for some $\alpha \in (0, 1)$ we get a variance upper bound that equals $\tfrac {1 - \alpha} {\alpha} \bar m$ for small values of $\rho (\bar e)$. The per-step complexity of approximately computing the likelihood and its gradient in this case is  both higher than computing the LL for CIGAM\footnote{Briefly, for a $k$-uniform hypergraph, computing the positive part of the LL cost $O(km)$, and the negative part of the LL costs (on expectation over the draws) $O \left ( k \binom n k \log (\bar m / (\bar m - |\mathcal B|) \right )$.}. Logistic-CP can also be augmented with the use of features and a learnable map $\psi$ with parameters $\nu$ such that $z_i = \psi ( x_i | \nu)$. In this model, the core nodes are nodes with $z_i \ge 0$ and the periphery nodes are nodes with $z_i < 0$ respectively. 

We also use the concurrent and independently developed method of \cite{tudisco2022core} -- HyperNSM -- where a permutation $\pi$ of the nodes is generated and then hyperedges are independently generated with probability $\sigma \left ( \xi(e) M_a(\{ a_u \}_{u \in e}) \right )$, where $M_a$ denotes the generalized (H\"older) $a$-mean. $\xi(e)$ is a weight function (e.g. $\xi(e) = 1 / |e|$) and $a_u = 1 - \tfrac {\pi_u} n$. The goal of HyperNSM is to recover the optimal permutation $\pi^*$ of the nodes through a fixed-point iteration scheme.  To calculate the LL of HyperNSM we use negative sampling such as in the case of Logistic-CP.

Tab.~\ref{tab:hypergraph_experiments} shows the results of comparing CIGAM with Logistic-CP, and HyperNSM. As in Section 4.2 of \cite{jia2019random}, we compare the optimized LL values of both CIGAM, Logistic-CP, and HyperNSM. We run experiments both using \emph{exogenous} ranks (directly from the features), as well as \emph{endogenous} (learnable) ranks via the mined features. We also report the learned core profile $c^*$ and the learned $\lambda^*$ for all datasets. The breakpoints are taken with step $0.5$. 
In almost all of the experiments, CIGAM has a \emph{substantially better} optimal LL than its competitors, which in many cases cannot scale (because GPUs run out of memory) to even moderate dataset sizes (denoted by $\dag$). In terms of learned parameters, we observe that most of the datasets are very sparse (very high values of $c^*$) in both the exogenous and the endogenous case. In App.~\ref{app:regularization}, we show how to add regularization (or priors) to the model. 

\noindent \textbf{Projected Hypergraph Experiments.} We run the same experiments as in Tab.~\ref{tab:hypergraph_experiments} but instead of the hypergraph data, we use the \emph{projected} graphs that result from replacing each hyperedge with a clique. We also use the model of \cite{tudisco2019nonlinear}, together with Logistic-CP and CIGAM. More specifically, the model of \cite{tudisco2019nonlinear}, which we call Logistic-TH, is a logistic-based model (the graph analogue of HyperNSM) that depends on finding a permutation $\pi$ of the nodes. In this model, and edge $(u, v)$ is generated independently with probability $\sigma (M_a (a_u, a_v))$ where $a_u = 1 - \frac {\pi_u} n$ (resp. $a_v$). The authors devise an iterative method to optimize the LL of the corresponding generative model. The iterative method converges to a fixed point vector $x^*$, and the ordering of the elements of $x^*$ implies the optimal permutation $\pi^*$. Again, to calculate the LL after finding $\pi^*$ we use negative sampling.  

We report the experimental results in Tab.~\ref{tab:graph_experiments}. Again, we observe that CIGAM is able to find better fits than both Logistic-CP and Logistic-TH, while it is also able to scale more smoothly to the largest datasets. Furthermore, we observe that the values of $c^*$ that CIGAM finds compared to the hypergraph case are \emph{substantially smaller}. This can be attributed to the fact that the projected hypergraphs have a smaller possible number of edges (i.e. $\binom n 2$), and, thus, they are \emph{``denser''} than the hypergraph instances since hyperedges are projected on the same order. 

\noindent \textbf{Ranking Comparison.} In Fig.~\ref{fig:spearman_algorithms} we compare the node orderings produced by CIGAM and Logistic-CP (and Logistic-TH in the graph case) on threads-ask-ubuntu and threads-math-sx (LCC + 2-core). We observe that in threads-ask-ubuntu have uncorrelated rankings, whereas in the case of threads-math-sx the rankings are negatively correlated. So, in both cases, the two algorithms are producing different rankings. Finally, we observe that both Logistic-CP and CIGAM are uncorrelated with the rankings that Logistic-TH produces. This is indicative of the algorithms having a different way on ranking the nodes. Indeed, Logistic-CP creates edges based on the sum of the core scores of the nodes, CIGAM creates edges based on the nodes with the maximum and minimum ranks, while Logistic-TH uses the generalized mean to resolve a version of the problem in the spectrum ``between'' CIGAM and Logistic-CP.


\noindent \textbf{Applications \& Future Work.} 
Similarly to graphs (see e.g. \cite{papachristou2021sublinear}), hypergraphs with a small core can be utilized to speed up computational tasks ( community detection, clustering, embeddings, user modeling, fandom formation etc.) that can be performed within the \emph{sublinear core} and then the results can be aggregated to the periphery. 

Finally, it should be noted that CIGAM does not have control over motifs, and the experimental or theoretical study of the distributions of motifs is interesting future work. 

\medskip

\noindent \textbf{Acknowledgements.} \funding
\acknowledgements

\bibliographystyle{ACM-Reference-Format}
\bibliography{main}


\begin{thebibliography}{47}


\ifx \showCODEN    \undefined \def \showCODEN     #1{\unskip}     \fi
\ifx \showDOI      \undefined \def \showDOI       #1{#1}\fi
\ifx \showISBNx    \undefined \def \showISBNx     #1{\unskip}     \fi
\ifx \showISBNxiii \undefined \def \showISBNxiii  #1{\unskip}     \fi
\ifx \showISSN     \undefined \def \showISSN      #1{\unskip}     \fi
\ifx \showLCCN     \undefined \def \showLCCN      #1{\unskip}     \fi
\ifx \shownote     \undefined \def \shownote      #1{#1}          \fi
\ifx \showarticletitle \undefined \def \showarticletitle #1{#1}   \fi
\ifx \showURL      \undefined \def \showURL       {\relax}        \fi
\providecommand\bibfield[2]{#2}
\providecommand\bibinfo[2]{#2}
\providecommand\natexlab[1]{#1}
\providecommand\showeprint[2][]{arXiv:#2}

\bibitem[\protect\citeauthoryear{Amburg, Kleinberg, and Benson}{Amburg
  et~al\mbox{.}}{2021}]%
        {amburg2021planted}
\bibfield{author}{\bibinfo{person}{Ilya Amburg}, \bibinfo{person}{Jon
  Kleinberg}, {and} \bibinfo{person}{Austin~R Benson}.}
  \bibinfo{year}{2021}\natexlab{}.
\newblock \showarticletitle{Planted hitting set recovery in hypergraphs}.
\newblock \bibinfo{journal}{\emph{Journal of Physics: Complexity}}
  \bibinfo{volume}{2}, \bibinfo{number}{3} (\bibinfo{year}{2021}),
  \bibinfo{pages}{035004}.
\newblock


\bibitem[\protect\citeauthoryear{Benson}{Benson}{2019}]%
        {benson2019three}
\bibfield{author}{\bibinfo{person}{Austin~R Benson}.}
  \bibinfo{year}{2019}\natexlab{}.
\newblock \showarticletitle{Three hypergraph eigenvector centralities}.
\newblock \bibinfo{journal}{\emph{SIAM Journal on Mathematics of Data Science}}
  \bibinfo{volume}{1}, \bibinfo{number}{2} (\bibinfo{year}{2019}),
  \bibinfo{pages}{293--312}.
\newblock


\bibitem[\protect\citeauthoryear{Benson, Abebe, Schaub, Jadbabaie, and
  Kleinberg}{Benson et~al\mbox{.}}{2018}]%
        {benson2018simplicial}
\bibfield{author}{\bibinfo{person}{Austin~R Benson}, \bibinfo{person}{Rediet
  Abebe}, \bibinfo{person}{Michael~T Schaub}, \bibinfo{person}{Ali Jadbabaie},
  {and} \bibinfo{person}{Jon Kleinberg}.} \bibinfo{year}{2018}\natexlab{}.
\newblock \showarticletitle{Simplicial closure and higher-order link
  prediction}.
\newblock \bibinfo{journal}{\emph{Proceedings of the National Academy of
  Sciences}} \bibinfo{volume}{115}, \bibinfo{number}{48}
  (\bibinfo{year}{2018}), \bibinfo{pages}{E11221--E11230}.
\newblock


\bibitem[\protect\citeauthoryear{Benson and Kleinberg}{Benson and
  Kleinberg}{2018}]%
        {benson2018found}
\bibfield{author}{\bibinfo{person}{Austin~R Benson} {and} \bibinfo{person}{Jon
  Kleinberg}.} \bibinfo{year}{2018}\natexlab{}.
\newblock \showarticletitle{Found Graph Data and Planted Vertex Covers}.
\newblock \bibinfo{journal}{\emph{Advances in Neural Information Processing
  Systems}}  \bibinfo{volume}{31} (\bibinfo{year}{2018}),
  \bibinfo{pages}{1356--1367}.
\newblock


\bibitem[\protect\citeauthoryear{Bonato, Janssen, and Pra{\l}at}{Bonato
  et~al\mbox{.}}{2010}]%
        {bonato2010geometric}
\bibfield{author}{\bibinfo{person}{Anthony Bonato}, \bibinfo{person}{Jeannette
  Janssen}, {and} \bibinfo{person}{Pawel Pra{\l}at}.}
  \bibinfo{year}{2010}\natexlab{}.
\newblock \showarticletitle{The geometric protean model for on-line social
  networks}. In \bibinfo{booktitle}{\emph{International Workshop on Algorithms
  and Models for the Web-Graph}}. Springer, \bibinfo{pages}{110--121}.
\newblock


\bibitem[\protect\citeauthoryear{Bonato, Janssen, and Pra{\l}at}{Bonato
  et~al\mbox{.}}{2012}]%
        {bonato2012geometric}
\bibfield{author}{\bibinfo{person}{Anthony Bonato}, \bibinfo{person}{Jeannette
  Janssen}, {and} \bibinfo{person}{Pawe{\l} Pra{\l}at}.}
  \bibinfo{year}{2012}\natexlab{}.
\newblock \showarticletitle{Geometric protean graphs}.
\newblock \bibinfo{journal}{\emph{Internet Mathematics}} \bibinfo{volume}{8},
  \bibinfo{number}{1-2} (\bibinfo{year}{2012}), \bibinfo{pages}{2--28}.
\newblock


\bibitem[\protect\citeauthoryear{Bonato, Lozier, Mitsche,
  P{\'e}rez-Gim{\'e}nez, and Pra{\l}at}{Bonato et~al\mbox{.}}{2015}]%
        {bonato2015domination}
\bibfield{author}{\bibinfo{person}{Anthony Bonato}, \bibinfo{person}{Marc
  Lozier}, \bibinfo{person}{Dieter Mitsche}, \bibinfo{person}{Xavier
  P{\'e}rez-Gim{\'e}nez}, {and} \bibinfo{person}{Pawe{\l} Pra{\l}at}.}
  \bibinfo{year}{2015}\natexlab{}.
\newblock \showarticletitle{The domination number of on-line social networks
  and random geometric graphs}. In \bibinfo{booktitle}{\emph{International
  Conference on Theory and Applications of Models of Computation}}. Springer,
  \bibinfo{pages}{150--163}.
\newblock


\bibitem[\protect\citeauthoryear{Borgatti and Everett}{Borgatti and
  Everett}{2000}]%
        {borgatti2000models}
\bibfield{author}{\bibinfo{person}{Stephen~P Borgatti} {and}
  \bibinfo{person}{Martin~G Everett}.} \bibinfo{year}{2000}\natexlab{}.
\newblock \showarticletitle{Models of core/periphery structures}.
\newblock \bibinfo{journal}{\emph{Social networks}} \bibinfo{volume}{21},
  \bibinfo{number}{4} (\bibinfo{year}{2000}), \bibinfo{pages}{375--395}.
\newblock


\bibitem[\protect\citeauthoryear{Boyd, Fitzgerald, Mahutga, and Smith}{Boyd
  et~al\mbox{.}}{2010}]%
        {boyd2010computing}
\bibfield{author}{\bibinfo{person}{John~P Boyd}, \bibinfo{person}{William~J
  Fitzgerald}, \bibinfo{person}{Matthew~C Mahutga}, {and}
  \bibinfo{person}{David~A Smith}.} \bibinfo{year}{2010}\natexlab{}.
\newblock \showarticletitle{Computing continuous core/periphery structures for
  social relations data with MINRES/SVD}.
\newblock \bibinfo{journal}{\emph{Social Networks}} \bibinfo{volume}{32},
  \bibinfo{number}{2} (\bibinfo{year}{2010}), \bibinfo{pages}{125--137}.
\newblock


\bibitem[\protect\citeauthoryear{Clauset, Arbesman, and Larremore}{Clauset
  et~al\mbox{.}}{2015}]%
        {clauset2015systematic}
\bibfield{author}{\bibinfo{person}{Aaron Clauset}, \bibinfo{person}{Samuel
  Arbesman}, {and} \bibinfo{person}{Daniel~B Larremore}.}
  \bibinfo{year}{2015}\natexlab{}.
\newblock \showarticletitle{Systematic inequality and hierarchy in faculty
  hiring networks}.
\newblock \bibinfo{journal}{\emph{Science advances}} \bibinfo{volume}{1},
  \bibinfo{number}{1} (\bibinfo{year}{2015}), \bibinfo{pages}{e1400005}.
\newblock


\bibitem[\protect\citeauthoryear{Clauset, Shalizi, and Newman}{Clauset
  et~al\mbox{.}}{2009}]%
        {clauset2009power}
\bibfield{author}{\bibinfo{person}{Aaron Clauset},
  \bibinfo{person}{Cosma~Rohilla Shalizi}, {and} \bibinfo{person}{Mark~EJ
  Newman}.} \bibinfo{year}{2009}\natexlab{}.
\newblock \showarticletitle{Power-law distributions in empirical data}.
\newblock \bibinfo{journal}{\emph{SIAM review}} \bibinfo{volume}{51},
  \bibinfo{number}{4} (\bibinfo{year}{2009}), \bibinfo{pages}{661--703}.
\newblock


\bibitem[\protect\citeauthoryear{De~Nooy, Mrvar, and Batagelj}{De~Nooy
  et~al\mbox{.}}{2018}]%
        {de2018exploratory}
\bibfield{author}{\bibinfo{person}{Wouter De~Nooy}, \bibinfo{person}{Andrej
  Mrvar}, {and} \bibinfo{person}{Vladimir Batagelj}.}
  \bibinfo{year}{2018}\natexlab{}.
\newblock \bibinfo{booktitle}{\emph{Exploratory social network analysis with
  Pajek: Revised and expanded edition for updated software}}.
  Vol.~\bibinfo{volume}{46}.
\newblock \bibinfo{publisher}{Cambridge University Press}.
\newblock


\bibitem[\protect\citeauthoryear{Della~Rossa, Dercole, and
  Piccardi}{Della~Rossa et~al\mbox{.}}{2013}]%
        {della2013profiling}
\bibfield{author}{\bibinfo{person}{Fabio Della~Rossa}, \bibinfo{person}{Fabio
  Dercole}, {and} \bibinfo{person}{Carlo Piccardi}.}
  \bibinfo{year}{2013}\natexlab{}.
\newblock \showarticletitle{Profiling core-periphery network structure by
  random walkers}.
\newblock \bibinfo{journal}{\emph{Scientific reports}} \bibinfo{volume}{3},
  \bibinfo{number}{1} (\bibinfo{year}{2013}), \bibinfo{pages}{1--8}.
\newblock


\bibitem[\protect\citeauthoryear{Eikmeier, Ramani, and Gleich}{Eikmeier
  et~al\mbox{.}}{2018}]%
        {eikmeier2018hyperkron}
\bibfield{author}{\bibinfo{person}{Nicole Eikmeier}, \bibinfo{person}{Arjun~S
  Ramani}, {and} \bibinfo{person}{David Gleich}.}
  \bibinfo{year}{2018}\natexlab{}.
\newblock \showarticletitle{The hyperkron graph model for higher-order
  features}. In \bibinfo{booktitle}{\emph{2018 IEEE International Conference on
  Data Mining (ICDM)}}. IEEE, \bibinfo{pages}{941--946}.
\newblock


\bibitem[\protect\citeauthoryear{Elliott, Chiu, Bazzi, Reinert, and
  Cucuringu}{Elliott et~al\mbox{.}}{2020}]%
        {elliott2020core}
\bibfield{author}{\bibinfo{person}{Andrew Elliott}, \bibinfo{person}{Angus
  Chiu}, \bibinfo{person}{Marya Bazzi}, \bibinfo{person}{Gesine Reinert}, {and}
  \bibinfo{person}{Mihai Cucuringu}.} \bibinfo{year}{2020}\natexlab{}.
\newblock \showarticletitle{Core--periphery structure in directed networks}.
\newblock \bibinfo{journal}{\emph{Proceedings of the Royal Society A}}
  \bibinfo{volume}{476}, \bibinfo{number}{2241} (\bibinfo{year}{2020}),
  \bibinfo{pages}{20190783}.
\newblock


\bibitem[\protect\citeauthoryear{Gelman, Lee, and Guo}{Gelman
  et~al\mbox{.}}{2015}]%
        {gelman2015stan}
\bibfield{author}{\bibinfo{person}{Andrew Gelman}, \bibinfo{person}{Daniel
  Lee}, {and} \bibinfo{person}{Jiqiang Guo}.} \bibinfo{year}{2015}\natexlab{}.
\newblock \showarticletitle{Stan: A probabilistic programming language for
  Bayesian inference and optimization}.
\newblock \bibinfo{journal}{\emph{Journal of Educational and Behavioral
  Statistics}} \bibinfo{volume}{40}, \bibinfo{number}{5}
  (\bibinfo{year}{2015}), \bibinfo{pages}{530--543}.
\newblock


\bibitem[\protect\citeauthoryear{Gousios}{Gousios}{2013}]%
        {Gousi13}
\bibfield{author}{\bibinfo{person}{Georgios Gousios}.}
  \bibinfo{year}{2013}\natexlab{}.
\newblock \showarticletitle{The GHTorrent dataset and tool suite}. In
  \bibinfo{booktitle}{\emph{Proceedings of the 10th Working Conference on
  Mining Software Repositories}} (San Francisco, CA, USA)
  \emph{(\bibinfo{series}{MSR '13})}. \bibinfo{publisher}{IEEE Press},
  \bibinfo{address}{Piscataway, NJ, USA}, \bibinfo{pages}{233--236}.
\newblock
\showISBNx{978-1-4673-2936-1}
\urldef\tempurl%
\url{http://dl.acm.org/citation.cfm?id=2487085.2487132}
\showURL{%
\tempurl}


\bibitem[\protect\citeauthoryear{Gousios and Spinellis}{Gousios and
  Spinellis}{2012}]%
        {gousios2012ghtorrent}
\bibfield{author}{\bibinfo{person}{Georgios Gousios} {and}
  \bibinfo{person}{Diomidis Spinellis}.} \bibinfo{year}{2012}\natexlab{}.
\newblock \showarticletitle{GHTorrent: GitHub's data from a firehose}. In
  \bibinfo{booktitle}{\emph{2012 9th IEEE Working Conference on Mining Software
  Repositories (MSR)}}. IEEE, \bibinfo{pages}{12--21}.
\newblock


\bibitem[\protect\citeauthoryear{Hoffman, Gelman, et~al\mbox{.}}{Hoffman
  et~al\mbox{.}}{2014}]%
        {hoffman2014no}
\bibfield{author}{\bibinfo{person}{Matthew~D Hoffman}, \bibinfo{person}{Andrew
  Gelman}, {et~al\mbox{.}}} \bibinfo{year}{2014}\natexlab{}.
\newblock \showarticletitle{The No-U-Turn sampler: adaptively setting path
  lengths in Hamiltonian Monte Carlo.}
\newblock \bibinfo{journal}{\emph{J. Mach. Learn. Res.}} \bibinfo{volume}{15},
  \bibinfo{number}{1} (\bibinfo{year}{2014}), \bibinfo{pages}{1593--1623}.
\newblock


\bibitem[\protect\citeauthoryear{Honaker, King, and Blackwell}{Honaker
  et~al\mbox{.}}{2011}]%
        {honaker2011amelia}
\bibfield{author}{\bibinfo{person}{James Honaker}, \bibinfo{person}{Gary King},
  {and} \bibinfo{person}{Matthew Blackwell}.} \bibinfo{year}{2011}\natexlab{}.
\newblock \showarticletitle{Amelia II: A program for missing data}.
\newblock \bibinfo{journal}{\emph{Journal of statistical software}}
  \bibinfo{volume}{45}, \bibinfo{number}{1} (\bibinfo{year}{2011}),
  \bibinfo{pages}{1--47}.
\newblock


\bibitem[\protect\citeauthoryear{Jia and Benson}{Jia and Benson}{2019}]%
        {jia2019random}
\bibfield{author}{\bibinfo{person}{Junteng Jia} {and} \bibinfo{person}{Austin~R
  Benson}.} \bibinfo{year}{2019}\natexlab{}.
\newblock \showarticletitle{Random spatial network models for core-periphery
  structure}. In \bibinfo{booktitle}{\emph{Proceedings of the Twelfth ACM
  International Conference on Web Search and Data Mining}}.
  \bibinfo{pages}{366--374}.
\newblock


\bibitem[\protect\citeauthoryear{Kaiser and Hilgetag}{Kaiser and
  Hilgetag}{2006}]%
        {Kaiser-2006-placement}
\bibfield{author}{\bibinfo{person}{Marcus Kaiser} {and}
  \bibinfo{person}{Claus~C. Hilgetag}.} \bibinfo{year}{2006}\natexlab{}.
\newblock \showarticletitle{Nonoptimal Component Placement, but Short
  Processing Paths, due to Long-Distance Projections in Neural Systems}.
\newblock \bibinfo{journal}{\emph{{PLoS} Computational Biology}}
  \bibinfo{volume}{2}, \bibinfo{number}{7} (\bibinfo{year}{2006}),
  \bibinfo{pages}{e95}.
\newblock
\urldef\tempurl%
\url{https://doi.org/10.1371/journal.pcbi.0020095}
\showDOI{\tempurl}


\bibitem[\protect\citeauthoryear{Kleinberg}{Kleinberg}{2002}]%
        {kleinberg2002small}
\bibfield{author}{\bibinfo{person}{Jon~M Kleinberg}.}
  \bibinfo{year}{2002}\natexlab{}.
\newblock \showarticletitle{Small-world phenomena and the dynamics of
  information}. In \bibinfo{booktitle}{\emph{Advances in neural information
  processing systems}}. \bibinfo{pages}{431--438}.
\newblock


\bibitem[\protect\citeauthoryear{Leskovec, Chakrabarti, Kleinberg, Faloutsos,
  and Ghahramani}{Leskovec et~al\mbox{.}}{2010}]%
        {leskovec2010kronecker}
\bibfield{author}{\bibinfo{person}{Jure Leskovec}, \bibinfo{person}{Deepayan
  Chakrabarti}, \bibinfo{person}{Jon Kleinberg}, \bibinfo{person}{Christos
  Faloutsos}, {and} \bibinfo{person}{Zoubin Ghahramani}.}
  \bibinfo{year}{2010}\natexlab{}.
\newblock \showarticletitle{Kronecker graphs: an approach to modeling
  networks.}
\newblock \bibinfo{journal}{\emph{Journal of Machine Learning Research}}
  \bibinfo{volume}{11}, \bibinfo{number}{2} (\bibinfo{year}{2010}).
\newblock


\bibitem[\protect\citeauthoryear{Leskovec, Kleinberg, and Faloutsos}{Leskovec
  et~al\mbox{.}}{2007}]%
        {leskovec2007graph}
\bibfield{author}{\bibinfo{person}{Jure Leskovec}, \bibinfo{person}{Jon
  Kleinberg}, {and} \bibinfo{person}{Christos Faloutsos}.}
  \bibinfo{year}{2007}\natexlab{}.
\newblock \showarticletitle{Graph evolution: Densification and shrinking
  diameters}.
\newblock \bibinfo{journal}{\emph{ACM transactions on Knowledge Discovery from
  Data (TKDD)}} \bibinfo{volume}{1}, \bibinfo{number}{1}
  (\bibinfo{year}{2007}), \bibinfo{pages}{2--es}.
\newblock


\bibitem[\protect\citeauthoryear{Menczer}{Menczer}{2002}]%
        {menczer2002growing}
\bibfield{author}{\bibinfo{person}{Filippo Menczer}.}
  \bibinfo{year}{2002}\natexlab{}.
\newblock \showarticletitle{Growing and navigating the small world web by local
  content}.
\newblock \bibinfo{journal}{\emph{Proceedings of the National Academy of
  Sciences}} \bibinfo{volume}{99}, \bibinfo{number}{22} (\bibinfo{year}{2002}),
  \bibinfo{pages}{14014--14019}.
\newblock


\bibitem[\protect\citeauthoryear{Nacher and Akutsu}{Nacher and Akutsu}{2012}]%
        {nacher2012dominating}
\bibfield{author}{\bibinfo{person}{Jose~C Nacher} {and}
  \bibinfo{person}{Tatsuya Akutsu}.} \bibinfo{year}{2012}\natexlab{}.
\newblock \showarticletitle{Dominating scale-free networks with variable
  scaling exponent: heterogeneous networks are not difficult to control}.
\newblock \bibinfo{journal}{\emph{New Journal of Physics}}
  \bibinfo{volume}{14}, \bibinfo{number}{7} (\bibinfo{year}{2012}),
  \bibinfo{pages}{073005}.
\newblock


\bibitem[\protect\citeauthoryear{Nacher and Akutsu}{Nacher and Akutsu}{2013}]%
        {nacher2013structural}
\bibfield{author}{\bibinfo{person}{Jose~C Nacher} {and}
  \bibinfo{person}{Tatsuya Akutsu}.} \bibinfo{year}{2013}\natexlab{}.
\newblock \showarticletitle{Structural controllability of unidirectional
  bipartite networks}.
\newblock \bibinfo{journal}{\emph{Scientific reports}} \bibinfo{volume}{3},
  \bibinfo{number}{1} (\bibinfo{year}{2013}), \bibinfo{pages}{1--8}.
\newblock


\bibitem[\protect\citeauthoryear{Nemeth and Smith}{Nemeth and Smith}{1985}]%
        {nemeth1985international}
\bibfield{author}{\bibinfo{person}{Roger~J Nemeth} {and}
  \bibinfo{person}{David~A Smith}.} \bibinfo{year}{1985}\natexlab{}.
\newblock \showarticletitle{International trade and world-system structure: A
  multiple network analysis}.
\newblock \bibinfo{journal}{\emph{Review (Fernand Braudel Center)}}
  \bibinfo{volume}{8}, \bibinfo{number}{4} (\bibinfo{year}{1985}),
  \bibinfo{pages}{517--560}.
\newblock


\bibitem[\protect\citeauthoryear{Newman et~al\mbox{.}}{Newman
  et~al\mbox{.}}{2003}]%
        {newman2003random}
\bibfield{author}{\bibinfo{person}{Mark~EJ Newman} {et~al\mbox{.}}}
  \bibinfo{year}{2003}\natexlab{}.
\newblock \showarticletitle{Random graphs as models of networks}.
\newblock \bibinfo{journal}{\emph{Handbook of graphs and networks}}
  \bibinfo{volume}{1} (\bibinfo{year}{2003}), \bibinfo{pages}{35--68}.
\newblock


\bibitem[\protect\citeauthoryear{Papachristou}{Papachristou}{2021}]%
        {papachristou2021sublinear}
\bibfield{author}{\bibinfo{person}{Marios Papachristou}.}
  \bibinfo{year}{2021}\natexlab{}.
\newblock \showarticletitle{Sublinear Domination and Core-Periphery Networks}.
\newblock \bibinfo{journal}{\emph{Scientific Reports}}  \bibinfo{volume}{11}
  (\bibinfo{year}{2021}).
\newblock


\bibitem[\protect\citeauthoryear{Papachristou and Kleinberg}{Papachristou and
  Kleinberg}{2022a}]%
        {anonymous_2022_5965856}
\bibfield{author}{\bibinfo{person}{Marios Papachristou} {and}
  \bibinfo{person}{Jon Kleinberg}.} \bibinfo{year}{2022}\natexlab{a}.
\newblock \bibinfo{booktitle}{\emph{Code - Core-periphery Models for
  Hypergraphs}}.
\newblock
\urldef\tempurl%
\url{https://doi.org/10.5281/zenodo.5965856}
\showDOI{\tempurl}


\bibitem[\protect\citeauthoryear{Papachristou and Kleinberg}{Papachristou and
  Kleinberg}{2022b}]%
        {anonymous_2022_5943044}
\bibfield{author}{\bibinfo{person}{Marios Papachristou} {and}
  \bibinfo{person}{Jon Kleinberg}.} \bibinfo{year}{2022}\natexlab{b}.
\newblock \bibinfo{booktitle}{\emph{Datasets - Core-periphery Models for
  Hypergraphs}}.
\newblock
\urldef\tempurl%
\url{https://doi.org/10.5281/zenodo.5943044}
\showDOI{\tempurl}


\bibitem[\protect\citeauthoryear{Ramani, Eikmeier, and Gleich}{Ramani
  et~al\mbox{.}}{2019}]%
        {ramani2019coin}
\bibfield{author}{\bibinfo{person}{Arjun~S Ramani}, \bibinfo{person}{Nicole
  Eikmeier}, {and} \bibinfo{person}{David~F Gleich}.}
  \bibinfo{year}{2019}\natexlab{}.
\newblock \showarticletitle{Coin-flipping, ball-dropping, and grass-hopping for
  generating random graphs from matrices of edge probabilities}.
\newblock \bibinfo{journal}{\emph{SIAM Rev.}} \bibinfo{volume}{61},
  \bibinfo{number}{3} (\bibinfo{year}{2019}), \bibinfo{pages}{549--595}.
\newblock


\bibitem[\protect\citeauthoryear{Rombach, Porter, Fowler, and Mucha}{Rombach
  et~al\mbox{.}}{2017}]%
        {rombach2017core}
\bibfield{author}{\bibinfo{person}{Puck Rombach}, \bibinfo{person}{Mason~A
  Porter}, \bibinfo{person}{James~H Fowler}, {and} \bibinfo{person}{Peter~J
  Mucha}.} \bibinfo{year}{2017}\natexlab{}.
\newblock \showarticletitle{Core-periphery structure in networks (revisited)}.
\newblock \bibinfo{journal}{\emph{SIAM review}} \bibinfo{volume}{59},
  \bibinfo{number}{3} (\bibinfo{year}{2017}), \bibinfo{pages}{619--646}.
\newblock


\bibitem[\protect\citeauthoryear{Sinha, Shen, Song, Ma, Eide, Hsu, and
  Wang}{Sinha et~al\mbox{.}}{2015}]%
        {sinha2015overview}
\bibfield{author}{\bibinfo{person}{Arnab Sinha}, \bibinfo{person}{Zhihong
  Shen}, \bibinfo{person}{Yang Song}, \bibinfo{person}{Hao Ma},
  \bibinfo{person}{Darrin Eide}, \bibinfo{person}{Bo-June Hsu}, {and}
  \bibinfo{person}{Kuansan Wang}.} \bibinfo{year}{2015}\natexlab{}.
\newblock \showarticletitle{An overview of microsoft academic service (mas) and
  applications}. In \bibinfo{booktitle}{\emph{Proceedings of the 24th
  international conference on world wide web}}. \bibinfo{pages}{243--246}.
\newblock


\bibitem[\protect\citeauthoryear{Snyder and Kick}{Snyder and Kick}{1979}]%
        {snyder1979structural}
\bibfield{author}{\bibinfo{person}{David Snyder} {and}
  \bibinfo{person}{Edward~L Kick}.} \bibinfo{year}{1979}\natexlab{}.
\newblock \showarticletitle{Structural position in the world system and
  economic growth, 1955-1970: A multiple-network analysis of transnational
  interactions}.
\newblock \bibinfo{journal}{\emph{American journal of Sociology}}
  \bibinfo{volume}{84}, \bibinfo{number}{5} (\bibinfo{year}{1979}),
  \bibinfo{pages}{1096--1126}.
\newblock


\bibitem[\protect\citeauthoryear{Stasi, Sadeghi, Rinaldo, Petrovi{\'c}, and
  Fienberg}{Stasi et~al\mbox{.}}{2014}]%
        {stasi2014beta}
\bibfield{author}{\bibinfo{person}{Despina Stasi}, \bibinfo{person}{Kayvan
  Sadeghi}, \bibinfo{person}{Alessandro Rinaldo}, \bibinfo{person}{Sonja
  Petrovi{\'c}}, {and} \bibinfo{person}{Stephen~E Fienberg}.}
  \bibinfo{year}{2014}\natexlab{}.
\newblock \showarticletitle{$beta$-models for random hypergraphs with a given
  degree sequence}.
\newblock \bibinfo{journal}{\emph{arXiv preprint arXiv:1407.1004}}
  (\bibinfo{year}{2014}).
\newblock


\bibitem[\protect\citeauthoryear{Tang, Zhang, Yao, Li, Zhang, and Su}{Tang
  et~al\mbox{.}}{2008}]%
        {tang2008arnetminer}
\bibfield{author}{\bibinfo{person}{Jie Tang}, \bibinfo{person}{Jing Zhang},
  \bibinfo{person}{Limin Yao}, \bibinfo{person}{Juanzi Li}, \bibinfo{person}{Li
  Zhang}, {and} \bibinfo{person}{Zhong Su}.} \bibinfo{year}{2008}\natexlab{}.
\newblock \showarticletitle{Arnetminer: extraction and mining of academic
  social networks}. In \bibinfo{booktitle}{\emph{Proceedings of the 14th ACM
  SIGKDD international conference on Knowledge discovery and data mining}}.
  \bibinfo{pages}{990--998}.
\newblock


\bibitem[\protect\citeauthoryear{Thorndike}{Thorndike}{1953}]%
        {thorndike1953belongs}
\bibfield{author}{\bibinfo{person}{Robert~L Thorndike}.}
  \bibinfo{year}{1953}\natexlab{}.
\newblock \showarticletitle{Who belongs in the family?}
\newblock \bibinfo{journal}{\emph{Psychometrika}} \bibinfo{volume}{18},
  \bibinfo{number}{4} (\bibinfo{year}{1953}), \bibinfo{pages}{267--276}.
\newblock


\bibitem[\protect\citeauthoryear{Tudisco and Higham}{Tudisco and
  Higham}{2019a}]%
        {tudisco2019fast}
\bibfield{author}{\bibinfo{person}{Francesco Tudisco} {and}
  \bibinfo{person}{Desmond~J Higham}.} \bibinfo{year}{2019}\natexlab{a}.
\newblock \showarticletitle{A fast and robust kernel optimization method for
  core--periphery detection in directed and weighted graphs}.
\newblock \bibinfo{journal}{\emph{Applied Network Science}}
  \bibinfo{volume}{4}, \bibinfo{number}{1} (\bibinfo{year}{2019}),
  \bibinfo{pages}{1--13}.
\newblock


\bibitem[\protect\citeauthoryear{Tudisco and Higham}{Tudisco and
  Higham}{2019b}]%
        {tudisco2019nonlinear}
\bibfield{author}{\bibinfo{person}{Francesco Tudisco} {and}
  \bibinfo{person}{Desmond~J Higham}.} \bibinfo{year}{2019}\natexlab{b}.
\newblock \showarticletitle{A nonlinear spectral method for core--periphery
  detection in networks}.
\newblock \bibinfo{journal}{\emph{SIAM Journal on Mathematics of Data Science}}
  \bibinfo{volume}{1}, \bibinfo{number}{2} (\bibinfo{year}{2019}),
  \bibinfo{pages}{269--292}.
\newblock


\bibitem[\protect\citeauthoryear{Tudisco and Higham}{Tudisco and
  Higham}{2022}]%
        {tudisco2022core}
\bibfield{author}{\bibinfo{person}{Francesco Tudisco} {and}
  \bibinfo{person}{Desmond~J Higham}.} \bibinfo{year}{2022}\natexlab{}.
\newblock \showarticletitle{Core-periphery detection in hypergraphs}.
\newblock \bibinfo{journal}{\emph{arXiv preprint arXiv:2202.12769}}
  (\bibinfo{year}{2022}).
\newblock


\bibitem[\protect\citeauthoryear{Wahlstr{\"o}m, Skog, La~Rosa, H{\"a}ndel, and
  Nehorai}{Wahlstr{\"o}m et~al\mbox{.}}{2016}]%
        {wahlstrom2016beta}
\bibfield{author}{\bibinfo{person}{Johan Wahlstr{\"o}m}, \bibinfo{person}{Isaac
  Skog}, \bibinfo{person}{Patricio~S La~Rosa}, \bibinfo{person}{Peter
  H{\"a}ndel}, {and} \bibinfo{person}{Arye Nehorai}.}
  \bibinfo{year}{2016}\natexlab{}.
\newblock \showarticletitle{The $beta$-model for Random Graphs---Regression,
  Cram\'er-Rao Bounds, and Hypothesis Testing}.
\newblock \bibinfo{journal}{\emph{arXiv preprint arXiv:1611.05699}}
  (\bibinfo{year}{2016}).
\newblock


\bibitem[\protect\citeauthoryear{Wallerstein}{Wallerstein}{1987}]%
        {wallerstein1987world}
\bibfield{author}{\bibinfo{person}{Immanuel Wallerstein}.}
  \bibinfo{year}{1987}\natexlab{}.
\newblock \showarticletitle{World-systems analysis}.
\newblock \bibinfo{journal}{\emph{Social theory today}}  \bibinfo{volume}{3}
  (\bibinfo{year}{1987}).
\newblock


\bibitem[\protect\citeauthoryear{Watts, Dodds, and Newman}{Watts
  et~al\mbox{.}}{2002}]%
        {watts2002identity}
\bibfield{author}{\bibinfo{person}{Duncan~J Watts},
  \bibinfo{person}{Peter~Sheridan Dodds}, {and} \bibinfo{person}{Mark~EJ
  Newman}.} \bibinfo{year}{2002}\natexlab{}.
\newblock \showarticletitle{Identity and search in social networks}.
\newblock \bibinfo{journal}{\emph{science}} \bibinfo{volume}{296},
  \bibinfo{number}{5571} (\bibinfo{year}{2002}), \bibinfo{pages}{1302--1305}.
\newblock


\bibitem[\protect\citeauthoryear{Zhang, Martin, and Newman}{Zhang
  et~al\mbox{.}}{2015}]%
        {zhang2015identification}
\bibfield{author}{\bibinfo{person}{Xiao Zhang}, \bibinfo{person}{Travis
  Martin}, {and} \bibinfo{person}{Mark~EJ Newman}.}
  \bibinfo{year}{2015}\natexlab{}.
\newblock \showarticletitle{Identification of core-periphery structure in
  networks}.
\newblock \bibinfo{journal}{\emph{Physical Review E}} \bibinfo{volume}{91},
  \bibinfo{number}{3} (\bibinfo{year}{2015}), \bibinfo{pages}{032803}.
\newblock


\end{thebibliography}

\newpage

\appendix

\section{Core Structure} \label{app:core}

\subsection{Proof of Theorem 1}

\textbf{Coupling construction.} Let a CIGAM model $G_1$ with parameters $\lambda$ and $1 < c_1 \le c_2 \dots \le c_L < e^\lambda$ be given and let a CIGAM model $G_2$ have parameters $\lambda$ and one layer with value $c_L$. We construct the coupling $\nu$ as follows: We first sample the rank vector $r$ (common for both $G_1$ and $G_2$) and then construct the hyperedges as follows: (i) If a hyperedge appears on $G_2$ then with probability 1 is appears on $G_1$, and (ii) if a hyperedge does does not appear on $G_2$ then it appears on $G_1$ with probability $\frac {f_1(e) - f_2(e)} {1 - f_2(e)}$. We can easily show that the marginals satisfy $\Pr [e \in E(G_1) | r] = f_2(e) \cdot 1 + (1 - f_2(e)) \frac {f_1(e) - f_2(e)} {1 - f_2(e)} = f_1(e)$ and $\Pr [e \in E(G_2) | r] = f_2(e)$. Finally, we integrate over $r$ to get that $\Pr [e \in E(G_1)] = f_1(e)$ and $\Pr [e \in E(G_2)] = f_2(e)$. Therefore $\nu$ is a valid coupling. Under $\nu$ we have that always $G_2 \subseteq G_1$. Therefore, it suffices to prove the Theorem for $G_2$ to get a result that holds for $G_1$. 

\noindent \textbf{Core size.} We prove the statement in the case that $G_1$ is $k$-uniform. Since by the coupling construction $G_2 \subseteq G_1$, proving a statement for the size of the core on $G_2$ will also hold for $G_1$ since a dominating set in $G_2$ is a dominating set in $G_1$. Let $t \in [0, 1]$ be a threshold value to be determined later. Define $$N_k(t) = \sum_{(j_1,\dots j_{k - 1}) \in \binom {[n]} {k - 1}} \one \{ \max \{ r(j_1), \\ \dots, r(j_{k - 1}) \} \ge t \}$$ be the number of nodes with ranks at least $t$. Note that by a simple combinatorial argument 

\begin{equation*}
\begin{split}
    N_k(t) & = \binom {n} {k - 1} - \left | J \in \binom {[n]} {k - 1} : \forall j \in J, r_j < t \right | \\ & = \binom {n} {k - 1} - \binom {n - N_2(t)} {k - 1}
\end{split}
\end{equation*}

where $N_2(t) \sim \mathrm{Bin} (n, 1 - F(t))$ is the number of nodes with $r_j \ge t$ and $\binom x y = \frac {\Gamma(x + 1)} {\Gamma(y + 1) \Gamma(x - y + 1)}$ is the generalized biniomial coefficient. The function $g(\nu) = \binom n {k - 1} - \binom {n - \nu} {k - 1}$ is strictly increasing for $0 \le \nu \le n$. Therefore, we can directly devise a concentration bound for $N_k(t)$ via concentration bounds for $N_2(t)$. Indeed, by the Chernoff bound 

\begin{align*}
\Pr \bigg [ N_k(t) \ge \binom n {k - 1} - \binom { n - \ev {} {N_2(t)} + \sqrt {n \log n / 2} } {k - 1} \bigg ] \\ = \Pr [N_2(t) \ge \ev {} {N_2(t)} - \sqrt {n \log n / 2}] \ge 1 - \frac 1 n
\end{align*} 

as long as $t \le F^{-1} (1 - \sqrt {\log n / (2n)})$. Note that the probability that a node $i \in [n]$ is not dominated by any core-hyperedge is given by 

\begin{align*}
    \Pr [\text {$i$ is not dominated by the core}] \\ = \prod_{J \in \binom {[n]} {k - 1}} \left ( 1 - c_L^{-2+\max_{j \in J \cup \{ i \}} r_j} \right )^{\one \{ \max_{j \in J} r_j \ge t \}} \\
     \le \prod_{J \in \binom {[n]} {k - 1}} \left ( 1 - c_L^{-2+t} \right )^{\one \{ \max_{j \in J} r_j \ge t \}} \\
    = \left ( 1 - c_L^{-2+t} \right )^{N_k(t)} \le \exp (-N_k(t) c_L^{-2+t}).
\end{align*}
    
If $N_k(t) \ge \frac {2 \log n} {c_L^{-2+t}}$ then by the union bound $\Pr [\exists i : \text {$i$ is not} \\ \text{dominated by the core} | N_k(t) \ge 2 \log n / c_L^{-2+t}] \le \frac 1 n$. Subsequently, the complementary event (i.e. $\forall i$, $i$ is  dominated by the core) happens with probability at least $1 - 1 / n$. We let $t$ to be such that

\begin{equation} \label{eq:threshold_core}
    \frac {2 \log n} {c_L^{-2+t}} = \binom {n} {k - 1} - \binom { nF(t) + \sqrt {n \log n / 2}  } {k - 1}, 
\end{equation}

in order for $\Pr [N_k(t) \ge 2 \log n / c_L^{-2+t}] \ge 1 - \tfrac 1 n$. Finally we have that given $t$ that satisfies \eqref{eq:threshold_core}

\begin{align*}
    \Pr \left [\text{Core at threshold $t$}\right ] & \ge \left ( 1 - \frac 1 n \right )^2 \ge 1 - \frac 2 n.
\end{align*}

\noindent \textbf{Existence and Uniqueness of threshold $t$.} We define 

\begin{equation}
    \Phi(t) = \frac {2 \log n} {c_L^{-2+t}} - \binom {n} {k - 1} + \binom { nF(t) + \sqrt {n \log n / 2}  } {k - 1}
\end{equation}

in the range $[0, t']$ where $t' = F^{-1} \left ( 1 - \sqrt {\frac {\log n} {2n}} \right )$ is the point where the difference of the binomial coefficients becomes 0. Note that $\Phi$ is continuous and differentiable in $[0, t']$ with derivative

\begin{align*}
    \Phi'(t) & = - 2 \log c_L \log n / c_L^{-2+t} + \binom {n F(t) + \sqrt {n \log n / 2}} {k - 1} n f(t) \psi(nF(t) \\ & + \sqrt {n \log n / 2}) \\ 
    & \ge -2 e^{2 \lambda} \lambda e^{-\lambda t} \log n + \binom {n F(t) + \sqrt {n \log n / 2}} {k - 1} n f(t) \psi(nF(t) \\ & + \sqrt {n \log n / 2}) \\
    & \ge n f(t) \left [ -2 e^{2 \lambda} + \binom {n F(t) + \sqrt {n \log n / 2}} {k - 1} \psi(nF(t) + \sqrt {n \log n / 2}) \right ] \\
    & \ge n f(t) \left [ -2 e^{2 \lambda} + \binom {\sqrt {n \log n / 2}} {k - 1} \psi(\sqrt {n \log n / 2}) \right ] \\
    & \ge n f(t) \left [ -2 e^{2 \lambda} + \sqrt {n \log n / 2} \psi(\sqrt {n \log n / 2}) \right ] \\
    & \ge n f(t) \left [ -2 e^{2 \lambda} + \sqrt {n \log n / 2} \left [ \frac 1 2 \log (n \log n / 2) - \frac 1 {\sqrt {n \log n / 2}}  \right ] \right ] \\
    & \ge n f(t) \left [ -3 e^{2 \lambda} + \sqrt {n / 8} \right ] > 0.
\end{align*}

for $\lambda < \frac {\ln (n / 72)} {4} \in o(\log n)$. $\psi(x) = \frac {\Gamma'(x)} {\Gamma(x)}$ is the digamma function. For the inequalities we have used the facts: (i) $c_L < e^\lambda$, (ii) $\log n \le n$, (iii) monotonicity of the Gamma and the digamma function for $t \ge 0$, (iv) $\binom n {k - 1} \ge n$ for $k < n - 1$, (v) $\psi(x) \ge \log x - \frac 1 x$, (vi) $n \ge 2$. Therefore $\Phi(t)$ is strictly increasing. Note that $\Phi(0) = 2 c_L^2 \log n - \binom {n} {k - 1} + \binom {\sqrt {n \log n}} {k - 1} < 0$ for large enough $n$ and since $c_L = o(n)$. Moreover note that $\Phi(t') = 2 \log n / c_L^{-2 + t'} > 0$. Therefore $\Phi(0) \Phi(t') < 0$. Thus by Bolzano's theorem and the fact that $\Phi'(t) > 0$ we get that there exists a unique threshold $t \in [0, t']$ such that $\Phi(t) = 0$.

\noindent \textbf{Upper Bound.} Note that the threshold is maximized when $k = 2$, i.e. in the graph case, and therefore the expected size of the core satisfies $\ev {} {\text{Core size}} \le n (1 - F(t(k =2))) = \sqrt {n \log n / 2} + 2 \log n / c_L^{-2+t} = \til O (\sqrt n)$ (see also Fig.~\ref{fig:core_size}).

\subsection{Empirical Core Thresholds}

\begin{figure}[h]
    \centering
    \includegraphics[width=0.9\columnwidth]{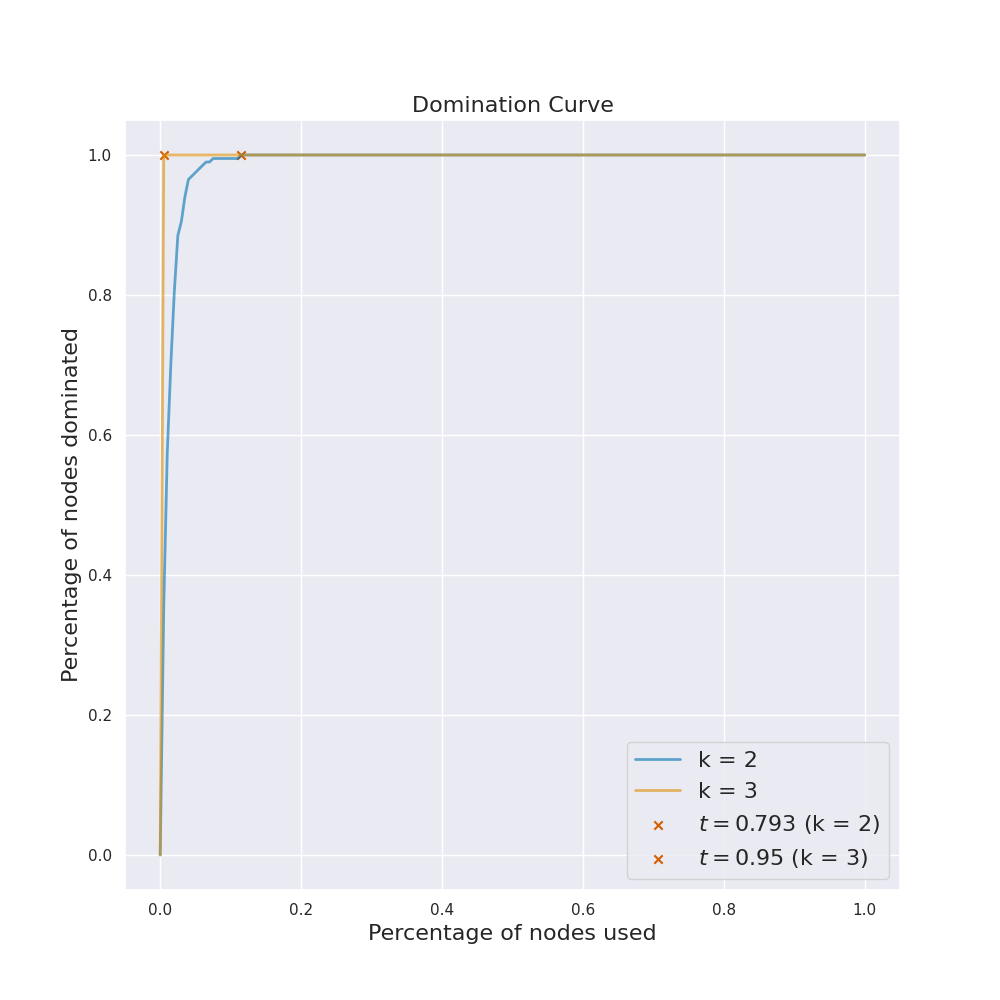}
    \vspace{-\baselineskip}
    \caption{Empirical core Threshold on generated instances for single layer model with $n = 200$, $b = 3.5$, $c_1 = 3$ for $k \in  \{ 2, 3 \}$. By \texttt{``\textcolor{orange}{x}''} we denote the sample core threshold values.}
    \label{fig:empirical_core_size}
    \vspace{-\baselineskip}
\end{figure}

\section{Sampling} \label{app:sampling}

\noindent \textbf{Uniformly Sampling from $\binom {[n]} k$.} To sample a $k$-tuple uniformly at random we use a rejection sampling algorithm: 

\manualnumber{1.} Initialize $S \gets \emptyset$. 

\manualnumber{2.} While $|S| \le k$ repeat: Sample $i$ uniformly from $[n]$, and if $i \notin S$, add $i$ to $S$.

\noindent \textbf{Ball-dropping (Single-Layer / $k$-uniform).} For each $i \in [n]$ we create a set $\mathcal B_i$ in which we sample $M_i$ edges by sampling an edge $e$ uniformly from $\binom {[n]} {k - 1}$ (with $i$ being the dominant node, and if $e \cup \{ i \}$ does not belong to $\mathcal B_i$ we add it to $\mathcal B_i$. 

\noindent \textbf{Sampling Negative Edges.} To sample edges from $e \in \bar E$ (i.e. $e \notin E$) we maintain a set $\mathcal B$ of certain size $b$ and while $| \mathcal B | \le b$ we sample uniformly an order $\{ K = k \}$ with probability $\frac {\bar m_k} {\bar m}$ and then we sample $\bar e$ from $\binom {[n]} {k}$  uniformly. If $\bar e \notin (E \cup \mathcal B)$ then we update $\mathcal B \gets \mathcal B \cup \{ \bar e \}$. 

\section{Implementations} \label{app:implementation} 

\noindent \textbf{Methods.} We implement point estimation and Bayesian inference algorithms as part of the evaluation process, which are available in the code supplement. Tab.~\ref{tab:complexity} shows the costs of fitting CIGAM on various occasions. 

\manualnumber{1.} \emph{Point Estimation (MLE/MAP).} We implement point estimation for the parameters $(\lambda, c)$ (or $(\lambda, c, \theta)$) of CIGAM with PyTorch using the log-barrier method. We use Stochastic Gradient Descent (SGD) to train the model for a certain number of epochs, until the learned parameters have converged to their final values. To avoid underflows, and because the resulting probabilities at each epoch are $\ll 1$ we use a smoothing parameter $\gamma$ (here we use $\gamma = 10^{-10}$). which we add to the corresponding probabilities to avert underflow. For MAP we add priors/regularization to $\lambda, c$ (see App.~\ref{app:regularization}). For Logistic-CP we use the following architecture to learn $z_i$'s: \texttt{Linear($d$, $d$) $\to$ ReLU $\to$ Linear($d$, $1$)}. 

\manualnumber{2.} \emph{Bayesian Inference (BI).} We implement the posterior sampling procedures with \emph{mc-stan} \cite{gelman2015stan} which offers highly efficient sampling using Hamiltonian Monte Carlo with No-U-Turn-Sampling (HMC-NUTS) \cite{hoffman2014no} and compiles a C++ model for BI. Note that $\mathcal K$ is a \emph{convex polytope}. Finally, we add priors (see App.~\ref{app:regularization}) on the parameters to form a posterior density to sample from. For BI, the stan model samples from the truncated density via defining the parameter \texttt{c0} as a \texttt{positive\_ordered} vector (which induces a log-barrier constraint on the log-posterior) and the parameter \texttt{c} is devised as \texttt{c0 + 1} in the \texttt{transformed parameters} block. The preprocessing takes place in the \texttt{transformed data} block, and the \texttt{model} block is responsible for the log-posterior oracle.

Tab.~\ref{tab:complexity} shows the complexity of the implemented methods.

\begin{table}
    \centering
    \footnotesize
    \caption{Experiments with Regularization $\alpha_c = 100, \alpha_\lambda = 1, \beta_\lambda = 2$ (LCC + 2-core) for the StackExchange datasets for SGD with step-size 0.001 and 10 epochs.}
    \vspace{-\baselineskip}
    \label{tab:regularization}
    \begin{tabular}{l|rr|rr}
        \toprule
        Dataset & $c^*$ & $\lambda^*$ & $c^*$ & $\lambda^*$ \\
        \midrule
        & \multicolumn{2}{c|}{Hypergraph} &  \multicolumn{2}{c}{Projected} \\
        \midrule 
        threads-ask-ubuntu & [11.1] & 1.3 & [11.1] & 1.3 \\
        threads-math-sx & [1522] & 1.9 & [44.1] & 1.7 \\
        threads-stack-overflow & [8.6e+11] & 10.5 & [2.1e+5] & 10.9 \\
        \bottomrule
    \end{tabular}
\end{table}

\section{Priors \& Regularization} \label{app:regularization}

For notational convenience, we refer to the priors using the variables defined in the stan model. For CIGAM we can use \emph{exponential} priors for \texttt{c} (or \texttt{c0} respectively), i.e. we can impose a penalty of the form $p(\texttt{c0}) \propto e^{-\alpha_c \sum_{i \in [L]} \texttt{c0}_i}$. Moreover, a stronger penalty can be applied in terms of a Pareto prior, i.e. $p(\texttt{c}) \propto e^{-\alpha_c \sum_{i \in [L]} \log (\texttt {c}_i)}$. For the rank parameter $\lambda$ we impose a $\texttt{Gamma}(\alpha_\lambda, \beta_\lambda)$ prior. 

For Logistic-CP we can use L2 regularization for $z_i$ which corresponds to a Gaussian prior $p(z) \propto e^{-\frac {\alpha_\theta} 2 \sum_{i \in [n]} z_i^2}$ to penalize large values of the core scores both in terms of core ($z_i \ge 0$) and periphery ($z_i < 0$) nodes. 

Tab.~\ref{tab:regularization} shows the learned parameters when regularization is applied.

\end{document}